\documentclass[fleqn,usenatbib]{mnras}

\usepackage{graphicx}	% Including figure files
\usepackage{amsmath}	% Advanced maths commands
\usepackage{amssymb}	% Extra maths symbols
\usepackage{url}
\usepackage{ragged2e}
\usepackage{wrapfig}
\usepackage{textcomp}
\usepackage{hyperref}
\usepackage{setspace}
\usepackage{fancyhdr}
\usepackage{pdfpages}
\usepackage{float}
\usepackage{epstopdf}
\usepackage[toc, page]{appendix}

\title[Genetic Algorithm on CMB]{Application of Genetic Algorithm to Estimate the Large Angular Scale Features of Cosmic Microwave Background}

\author[Parth Nayak and Rajib Saha]{
Parth Nayak$^{1}$
and Rajib Saha$^{1}$
\\
$^{1}$Department of Physics, Indian Institute of Science Education and Research (IISER) Bhopal, 462066, India\\
}

\date{Accepted XXX. Received YYY; in original form ZZZ}

\pubyear{2021}

\begin{document}
\label{firstpage}
\pagerange{\pageref{firstpage}--\pageref{lastpage}}
\maketitle

\begin{abstract}
	%\hl{
		Genetic Algorithm (GA) -- motivated by natural evolution -- is a robust method to estimate the global optimal solutions of problems involving multiple objective functions. In this article, for the first time, we apply GA to reconstruct the CMB temperature anisotropy map over large angular scales of the sky using (internal) linear combination (ILC) of the final-year WMAP and Planck satellite observations. To avoid getting trapped into a local minimum, we implement the GA with generous diversity in the populations by selecting pairs with diverse fitness coefficients and by introducing a small but significant amount of mutation of genes. We find that the new GA-ILC method produces a clean map which agrees very well with that obtained using the exact analytical expression of weights in ILC. By performing extensive Monte Carlo simulations of the CMB reconstruction using the GA-ILC algorithm, we find that residual foregrounds in the cleaned map are minimal and tend to occupy localized regions along the galactic plane. The CMB angular power spectrum shows no indication of any bias in the entire multipole range $2 \leq \ell \leq 32$ studied in this work. The error in the CMB angular power spectrum is also minimal and given entirely by the cosmic-variance-induced error. Our results agree well with those obtained by various other reconstruction methods by different research groups. This problem-independent robust GA-ILC method provides a flexible way towards the complex and challenging task of CMB component reconstruction in cosmology.
	%}
\end{abstract}


%%%%%%%%%%%%%%%%%%%%%%%%%%%%%%%%%%%%%%%%%%%%%%%%%%

%%%%%%%%%%%%%%%%% BODY OF PAPER %%%%%%%%%%%%%%%%%%

%-------------------------
%-------------------------
\section{Introduction}
\label{sec:intro}
%-------------------------
%-------------------------

\par{The cosmic microwave background (CMB) angular power spectrum is one of the most crucial probes of the cosmological parameters and the dynamical history
of the universe. The thermal radiation is observed by various satellite missions such as WMAP \citep{2013ApJS..208...20B} and Planck \citep{Planck2018}. This primordial signal, however, is largely contaminated by the emissions from within our Milky Way, close to the galactic equator. Small amount of extragalactic contamination is also seen spread through the whole sky. These sources of contamination are collectively called the “foreground” sources since we are concerned with the cosmic microwave \textit{background}. For instance, the hot interstellar dust inside the galaxy emits strongly at frequencies $\gtrsim$ 100 GHz \citep{10.1093/ptep/ptu065}. This is referred to as the dust emission and is one kind of foreground. Some other kinds of foreground are synchrotron and free-free emissions from galactic and extragalactic sources that dominate the total foreground component at the smaller frequencies ($\lesssim$ 30 GHz) \citep{2003ApJS..148....1B, 1999NewA....4..443B, 2003ApJS..148..135H}. Hence, an accurate estimation of the angular power spectrum of pure CMB from the foreground-contained observations is crucial for precision cosmology.}

Since the onset of the scientific observations of CMB, extensive research has been done to come up with ways of effective foreground removal and accurate CMB retrieval. As a consequence, several methods exist in literature for getting rid of the foreground. However, most of them make use of the underlying spectral and spatial models of the foreground~\citep{2003ApJS..148..135H, 2003ApJS..148...97B, 2007ApJS..170..288H}. For example, \cite{2003MNRAS.346.1089D} propose a method for multi-detector maps using spectral matching with a model. A Gibbs sampling approach \citep{4767596} is proposed and implemented to separate CMB and foreground components by \cite{Eriksen_2004, Eriksen_2007, 2008ApJ...672L..87E, 2008ApJ...676...10E}. \cite{COMMANDER2} extend this Bayesian method through Wiener filtering for multi-resolution CMB observations. A maximum likelihood approach to reconstruct CMB and foreground components using prior noise covariance information and foreground models is implemented by \cite{Erikson_2006, Gold_2011}. However, modeling all the foreground components, especially in some regions, e.g. the galactic center/ plane, is expectedly more complex and therefore the existing models and templates may not be complete and/or accurate enough. This may then propagate to the retrieved clean CMB map and the power spectrum. Hence, there may be uncertainties arising due to this incomplete information of the foreground frequency and spatial dependence in the foreground models~\citep{Dodelson_1997,1998ApJ...502....1T}. 

 A completely model-independent method is the internal linear combination (ILC) of multifrequency observations \citep{Tegmark_1996}. \cite{2012MNRAS.419.1163B, 2013MNRAS.435...18B} implement an algorithm using spherical needlets that enables localized cleaning under ILC in pixel and harmonic spaces. \cite{2011MNRAS.418..467R} explore a needlet ILC to separate millimetre astrophysical foregrounds using a multidimensional ILC filter. \cite{Sudevan_2018_g} explore the CMB posterior using a Gibbs sampling approach within a global ILC framework. \cite{Sudevan_2018} study the method in the wake of the theoretical CMB covariance information. Other notable studies of the ILC method include, but not limited to,~\cite{Tegmark_2003,2003ApJS..148...97B, Eriksen_2004a,Saha_2006,Saha_2008,Saha_2011}. Another, notable CMB retrieval method, besides the external model-based approaches and the ILC-based methods, is the \emph{internal} template-fitting approach using a wavelet decomposition as described by \cite{2008A&A...491..597L, 2012MNRAS.420.2162F}. We refer the reader, for a detailed account of the prominent foreground reconstruction methods in literature, to  \cite{Planck2018IV}.

Some of the previous works involving ILC suggest that, depending on the context of the underlying problem, an exact expression of ILC weights may be extremely complicated or impossible to find. For instance, \cite{Saha_2011} proposes to reconstruct CMB by numerically minimizing the kurtosis of the ILC map, since an analytical solution of optimal ILC weights is unfeasible in that case. Problems of this sort require novel, innovative solutions. In this work, we come up with such a solution.

We propose an expression-independent, numerical ILC technique for a robust and effective CMB foreground reconstruction. By expression-independent, we mean that this method does not require an analytical expression of the optimal ILC weight-vector. Depending upon the strategy of the ILC under consideration, our numerical method can be adjusted to optimize a cost function of the ILC framework. In this work, we apply the novel numerical method over the global ILC proposed by \cite{Sudevan_2018}, making use of the Planck mission’s theoretical CMB angular power spectrum \citep{Planck2018VI} in our computation of the reduced variance of the output map. We apply the genetic algorithm (GA) for estimating optimal ILC weights to produce a clean map. We call the method ``GA-ILC'' as shorthand.

Genetic algorithm (GA) was first proposed by John H. Holland \citep{10.1145/321127.321128} in the mid 1960s and was further extended by David Goldberg in the late 1980s \citep{Goldberg}. GA aims to find global optimal solution(s) to objective functions having a multimodal nature, i.e., it has multiple local optima in the domain of search. As the conventional optima-finding algorithms tend to fail in such cases, GA opens up a new window with its novel concept. The underlying principle of GA is Darwin's theory of ``natural selection''. The biological species evolve by this natural process. In a population, each individual has a different set of genetic characteristics. Thus, the whole population has a diverse gene pool. This community is subject to certain natural circumstances known as the environment. The process of natural selection asserts that the genes which lead to better traits for survival will more likely be passed on to the offspring by reproduction over the generations than those which do not. Occasional mutations may bring in new beneficial genes that were earlier not present. Hence, eventually the whole population will best adapt to its environment after several generations. Evolutionary algorithms such as GA employ this strategy to optimize a given system under a given environment (as determined by the cost function).

For our study of the method over the large angular scales, we make use of the low-resolution WMAP and Planck observations as input at HEALPix~\footnote{Hierarchical Equal Area isoLatitude Pixelation of sky, e.g., see~\cite{Gorski2005}} $N_{\text{side}} = 16$. We smooth the input maps by a Gaussian beam at 9\textdegree \hspace{1pt} FWHM. Presumably, this also reduces the pixel-uncorrelated noise level (which is dominant on smaller angular scales).

We perform detailed Monte Carlo simulations to ensure the statistical sanity of our method. We find that this novel GA-ILC method produces a clean map with a minimal residual foreground. The residuary contamination tends to occupy small localized regions close to the galactic plane. The angular power spectrum of the clean map contains no apparent bias in the multipole range of $2 \leq \ell \leq 32$. The reconstruction errors in the angular power spectrum are also minimal and conform to the cosmic variance-induced errors.

%%(Paper organization paragraph)
The organization of this paper is as follows. We describe the basic formalism of the method in section \ref{sec:formalism}. We elaborate on the exact implementation of this method in section \ref{sec:meth}. We present our trial GA implementation for validation in section~\ref{sec:valid}. We discuss the GA-ILC implemented with Monte Carlo simulations in section~\ref{sec:MC-simul}. We present the results obtained by using the techniques of this work on WMAP and Planck final-year data and discuss our findings in section \ref{sec:results}. We conclude our paper by outlining some key aspects of this work and its future prospects in section \ref{sec:conclusion}.\\

%-------------------------
%-------------------------
\section{Formalism}
\label{sec:formalism}
%-------------------------
%-------------------------

All observations of the microwave sky consist primarily of CMB and foreground. Various foreground emissions are not thermalized, hence, they do not have a blackbody spectrum unlike CMB, which follows the thermal blackbody distribution with great accuracy \citep{1994ApJ...420..439M}. Hence, across different frequency bands, the antenna temperature of the foreground patches will vary whereas that of CMB will be constant. ILC exploits this fact as a benefit of the multifrequency information.
	
Apart from foreground, detector-noise is also one of the components that contaminates the observed CMB signal. Although small amount of noise is present on all angular scales, it is dominant only on the smaller angular scales of the data which are irrelevant for this work. Furthermore, noise from any two frequency channels is uncorrelated \citep{2003ApJS..148..135H, 2007ApJS..170..288H, 2003ApJS..145..413J, 2007ApJS..170..263J}, more to the advantage of ILC.

Suppose, under the instrumental framework of a CMB experiment, there are $N$ channels (or bands) of different frequencies, each producing one temperature anisotropy map (hereinafter simply, ``map'') upon observation. In principle, such a map in the $i$th band can be expressed as the following, after accounting for the beam effect of the telescope.
%%%%
\begin{eqnarray}\label{eqn:map-in-principle}
	\Delta T_i (\mathbf{\hat{n}}) = \int \Big[ \Delta T_{\text{CMB}} (\mathbf{\hat{n}'}) + \Delta T_{\text{fg}, i} (\mathbf{\hat{n}'}) \Big] B_i(\mathbf{\hat{n}} \cdot \mathbf{\hat{n}'})d\mathbf{\hat{n}'} \nonumber \\ 
	+ \Delta T_{\text{noise}, i} (\mathbf{\hat{n}}),
\end{eqnarray}
where
$\mathbf{\hat{n}}$ is the unit vector of direction in the sky and $B_i$ is the beam function of the $i$th band. The detector-noise component is free of the beam effects as we can see. We work with the 9\textdegree beam-smoothed data so that all multifrequency maps will have the same beam resolution. This drops the $i$ -index from the beam function.
  
In practice, however, the observed maps are unavoidably pixelized. Thus, the map in the $i$th frequency channel is denoted by the vector $\Delta \tilde{\mathbf{T}}_i$ (in thermodynamic temperature) in the HEALPix pixelated format. The size of this vector is $p \equiv N_{\text{pix}} = 12N_{\text{side}}^2$. This map can be written as the sum of its components in the following way:
%%%%
\begin{eqnarray}\label{eqn:cmb_comp}
	\Delta \tilde{\mathbf{T}}_i = \Delta \tilde{\mathbf{T}}_{\text{CMB}} 
	+ \Delta \tilde{\mathbf{T}}_{\text{fg}, i}
	+ \Delta \tilde{\mathbf{T}}_{\text{noise}, i}.
\end{eqnarray} 
Here, the beam effect has already been encompassed into the CMB and foreground parts.

Since a map is defined over a sphere, it can be expanded in the so-called ``harmonic space'' as
%%%%
\begin{eqnarray}\label{eqn:harm-space}
    \Delta \tilde{T}_i(n) = \sum_{\ell m} a_{\ell m}^i Y_{\ell m} (n),
\end{eqnarray}
where
$n$ denotes the pixel index (corresponding to some direction vector $\mathbf{\hat{n}}$), $Y_{\ell m}$ are the standard spherical harmonics, and $a_{\ell m}^i$ are called the harmonic coefficients of the $i$th map. The multipole index $\ell$ is of special importance because it indicates the underlying angular scale ($\theta_{\ell} \sim 2\pi/\ell$). The cross-correlation of the harmonic coefficients of maps $i$ and $j$ is also called the cross power spectrum, defined by
%%%%%
\begin{eqnarray}\label{eqn:cross-cl-dfn}
    C_{\ell}^{ij} = \sum_{m=-\ell}^{\ell} \frac{a_{\ell m}^{i*} a_{\ell m}^{j}}{2\ell + 1}.
\end{eqnarray}
The auto-correlation of the harmonic coefficients of the $i$th map is called its angular power spectrum, $C_{\ell}^i$. The CMB angular power spectrum is the single, most important quantity we would like to accurately estimate by finding an optimal clean CMB map from multifrequency observations. \\

%-------------------------
\subsection{Internal Linear Combination}
\label{sub:ilc}
%-------------------------

\cite{Tegmark_1996} proposed a model-independent method to estimate the clean CMB signal from a multifrequency observation. ILC attempts to find the optimal estimate of CMB from multifrequency foreground-contaminated maps without using any external templates or models. In this technique, we start by writing a linear combination of $N$ multifrequency input maps to find a clean output map. This is simply
%%%%%
\begin{eqnarray}\label{eqn:ilc}
	\mathbf{P} \equiv \Delta \tilde{\mathbf{T}}_{\text{clean}} (\{w_i\}) = \sum_{i=1}^{N} w_i \Delta \tilde{\mathbf{T}}_i,
\end{eqnarray}
where
\{$w_i$\} are the coefficients, more commonly called the \textit{weights} given to all the input maps before summing them. Using Eqn.~\eqref{eqn:cmb_comp}, this yields
%%%%%%
\begin{eqnarray}\label{eqn:ilc-expn}
	\mathbf{P} = \bigg( \sum_{i=1}^{N} w_i \bigg) \Delta \tilde{\mathbf{T}}_{\text{CMB}} 
	+ \sum_{i=1}^{N} w_i \Delta \tilde{\mathbf{T}}_{\text{trash}, i}, 
\end{eqnarray}
where
$\Delta \tilde{\mathbf{T}}_{\text{trash}, i} = \Delta\tilde{\mathbf{T}}_{\text{fg}, i} + \Delta \tilde{\mathbf{T}}_{\text{noise}, i}$.
In order to preserve the norm of the clean CMB map, a constraint is imposed in the form
%%%%%
\begin{eqnarray}\label{eqn:ilc-constraint}
	\sum_{i=1}^{N} w_i = 1 \text{\hspace{12pt} or \hspace{12pt}}
	\mathbf{e}^{\text{T}}\mathbf{w} = 1,
\end{eqnarray}
where
$\mathbf{e} = (1, 1,\dots,1)^{\text{T}}$ is the column vector of size $N$ with all entries 1. In the usual ILC method, the best estimate of CMB is obtained by minimizing the variance of the clean map, $\mathbf{P}^{\text{T}} \mathbf{P}$, w.r.t. the choice of the weight vector, $\mathbf{w}$. This is a problem of multivariable constrained optimization and the analytical solution can be found by using the Lagrange's multipliers method (e.g., see \cite{Tegmark_1996, 2008PhRvD..78b3003S}) as,
%%%%%
\begin{eqnarray}\label{eqn:w-lm-vector}
	\mathbf{w}_{\text{usual}} = \frac{\mathbf{e}^{\text{T}} 
  \mathbf{C}^{-1}}{\mathbf{e}^{\text{T}} \mathbf{C}^{\text{-1}} \mathbf{e}},
\end{eqnarray}
or following summation notation,
\begin{eqnarray}\label{eqn:w-lm-summ}
	w_{\text{usual}, i} = \frac{\sum_{j=1}^{N}C^{-1}_{ij}}{\sum_{i',j'=1}^{N} \text{ } C^{-1}_{i'j'}},
\end{eqnarray}
where
$\mathbf{C}$ is the $N \times N$ square symmetric covariance matrix of the $N$ input maps. In case the covariance matrix is singular, it is possible to generalize the inverse as the Moore-Penrose pseudoinverse~\citep{Moore_1920,1956PCPS...52...17P} $\mathbf{C}^{-1} \mapsto \mathbf{C}^{\dagger}$. If 
$\mathbf{C}$ is not singular, $\mathbf{C}^{\dagger} = \mathbf{C}^{-1}$. Rewriting Eqn.~\eqref{eqn:w-lm-vector} 
in terms of the pseudoinverse,  
%%%%%%%%
\begin{eqnarray}\label{eqn:w-lm-pi}
	\mathbf{w}_{\text{usual}} = \frac{\mathbf{e}^{\text{T}} \mathbf{C}^{\dagger}}{\mathbf{e}^{\text{T}} \mathbf{C}^{\dagger} \mathbf{e}}.
\end{eqnarray}

\cite{Sudevan_2018} (SS, for short) proposed a new, global ILC method making use of the prior knowledge of the theoretical covariance of CMB. The method employs this additional information to assert that the covariance structure of the final clean map is consistent with the expectation of pure CMB. Instead of minimizing the variance of the clean map, this global ILC seeks to minimize the reduced variance, called $\sigma^2$, calculated as
%%%%%%%%%%%
\begin{eqnarray}\label{eqn:sigma-sq-SS}
\sigma^2 =  \mathbf{P}^{\text{T}}
            \mathcal{C}_{\text{th}}^{\dagger} \mathbf{P},    
\end{eqnarray}
where
$\mathcal{C}_{\text{th}}$ is the theoretical CMB \emph{pixel-pixel} covariance matrix of size $p\times p$ and $\dagger$ represents the pseudoinverse. From the definition of $\mathbf{P}$ in Eqn.~\eqref{eqn:ilc}, we can write
%%%%%%%%%%%%%%
\begin{eqnarray}\label{eqn:sigma-sq-A}
    \sigma^2 =  \mathbf{w}^{\text{T}} \mathbf{A w},
\end{eqnarray}
where $\mathbf{A}$ is an $N\times N$ square matrix with elements
%%%%%%%%%%%%%
\begin{eqnarray}\label{eqn:A-elem-dfn}
A_{ij} = (\Delta \tilde{\mathbf{T}}_i)^{\text{T}}
            \mathcal{C}_{\text{th}}^{\dagger} 
            \Delta \tilde{\mathbf{T}}_j.
\end{eqnarray}
Similarly to the usual ILC, using the Lagrange multipliers with the same constraint, the optimal weights under this global ILC can be found as
%%%%%%%%%%
\begin{eqnarray}\label{eqn:w-SS}
    \mathbf{w}_{\text{SS}} =     \frac{\mathbf{e}^{\text{T}} \mathbf{A}^{\dagger}}{\mathbf{e}^{\text{T}} \mathbf{A}^{\dagger} \mathbf{e}}.
\end{eqnarray}

As found by \cite{2018arXiv181008872S}, the matrix $\mathbf{A}$ can be calculated as
\begin{eqnarray}\label{eqn:A-matrix}
	A_{ij} = \sum_{\ell = 2}^{\ell_{\text{max}}} (2\ell + 1) \frac{C_{\ell}^{ij}}{C_{\ell, \text{th}}'},	
\end{eqnarray}
where
$C_{\ell, \text{th}}'$ is the theoretical power spectrum of CMB after accounting for the beam and pixel effects, \textit{i.e.} 
\begin{eqnarray}\label{eqn:beam-pixel}
	C_{\ell, \text{th}}' = C_{\ell, \text{th}} B_{\ell}^2 P_{\ell}^2,
\end{eqnarray}
where
$B_{\ell}$ are the Legendre transforms of the beam function and $P_{\ell}$ are the HEALPix pixel 
window functions. $\ell_{\text{max}}$ can be chosen according to the pixel angular resolution of 
the maps. In this work, we made use of the low resolution maps at $N_{\text{side}} = 16$, for 
which $\ell = 32$ is a sufficiently large multipole. The sum is taken starting from $\ell = 2$ 
because the monopole ($\ell = 0$) and dipole ($\ell = 1$) are uninteresting for cosmological studies, and hence, they are removed from the input maps prior to all the analyses.

(The reduced variance of some map may be calculated as
\begin{eqnarray}
	\sigma^2 = \sum_{\ell = 2}^{\ell_{\text{max}}} (2\ell + 1) \frac{C_{\ell}}{C_{\ell, \text{th}}'},	
\end{eqnarray}
where $C_{\ell}$ is the angular power spectrum computed from the given map.)\\

%-------------------------
\subsection{Genetic Algorithm}
\label{sub:GA_form}
%-------------------------
%%
\begin{figure}
	\begin{center}
		\includegraphics[width=0.9\columnwidth]{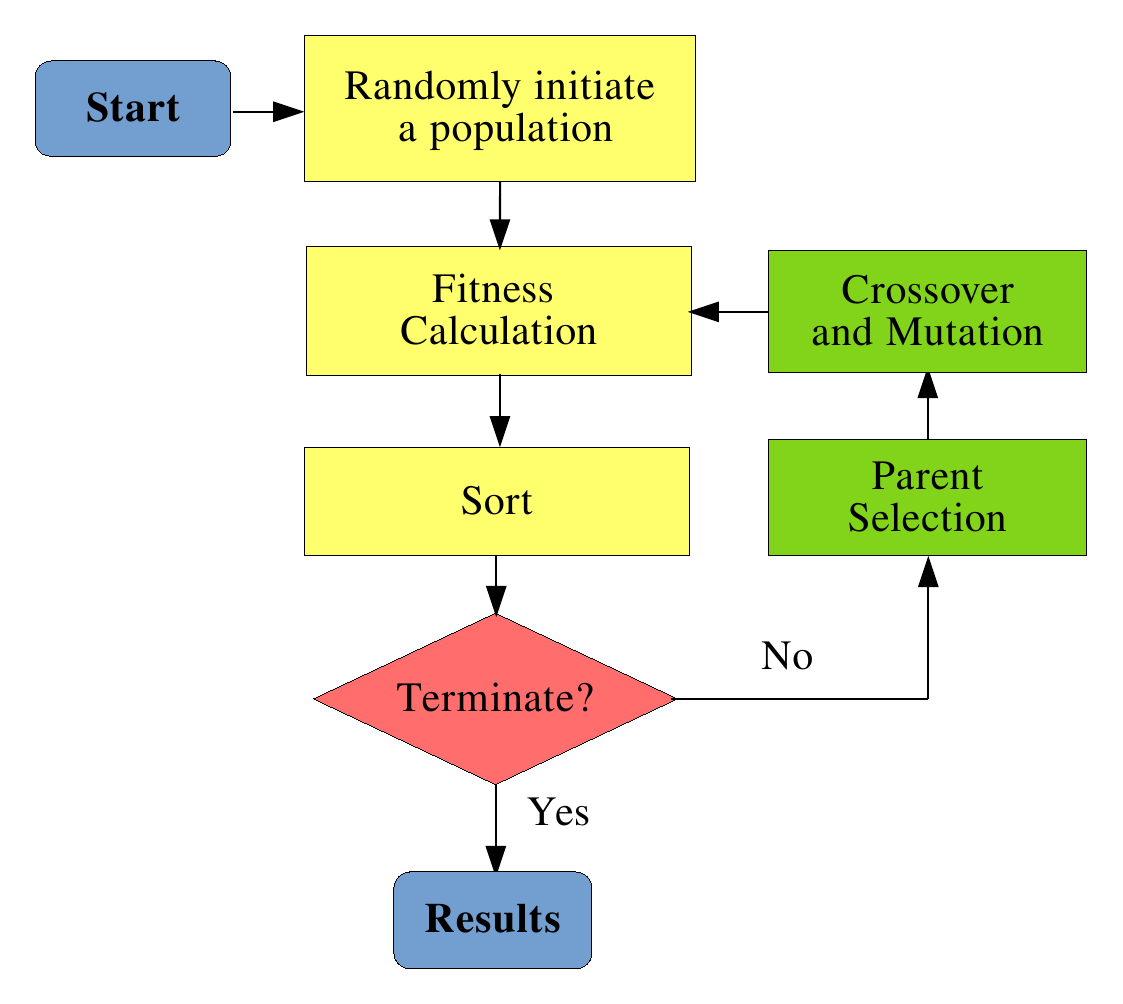}
	\end{center}
	\caption{The flowchart of a typical GA. The sorting is introduced in order to ensure the better readability and presentability of the solutions in a population.}\label{fig:GA-flow}
\end{figure}

 Evolution by natural selection is a stochastic process, and the origination and adaptation of species occur randomly over a long time. Since GA makes use of this concept as a backbone, it is obvious that the algorithm itself is stochastic in nature. (In fact, it is part of the reason behind the robustness of such algorithms.) Indeed, there exists no problem-independent framework of this evolutionary algorithm. A problem-specific investigation of a suitable GA architecture is inevitable. Moreover, the hyper-parameters of a GA do not follow any analytical expressions by which one could hope to determine optimal values of such factors.
 	
 An overall guideline to choose optimal hyper-parameters for a GA such as ours is as follows: Initially a number of test-runs are made choosing a variety of parameter values. This is repeated a few times observing the trends of the final results of the GA \emph{w.r.t.} the changes in individual parameter values. This quickly directs the programmer towards a set of optimized parameter values. We employ this method of establishing the hyper-parameters for our GA-ILC. We refer to this strategy henceforth as the \emph{hit-and-trial} method for our parameter-choosing. We further discuss it in subsection~\ref{sub:implmn}.

\begin{figure*}
	\includegraphics[width=0.26\linewidth]{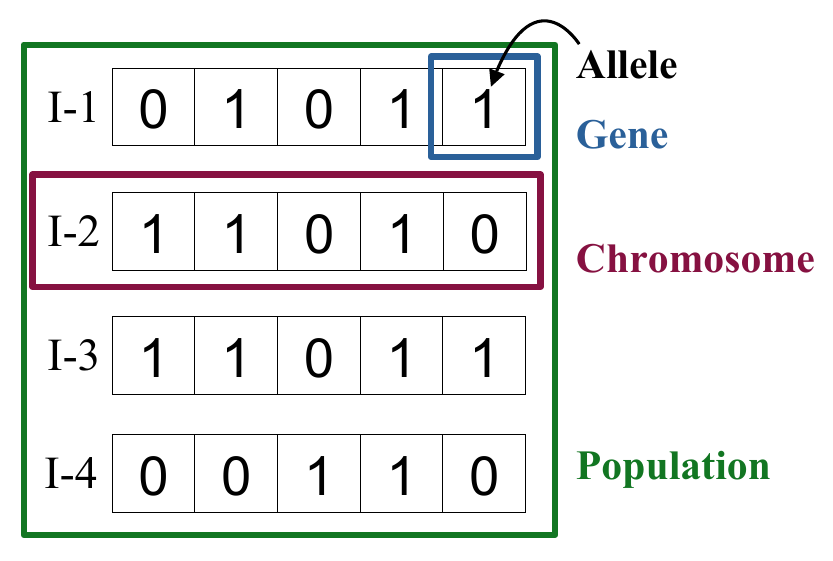}
		{\color{gray} \hspace{1pt}\rule{0.4pt}{90pt}\hspace{1pt}} 
	\includegraphics[width=0.2\linewidth]{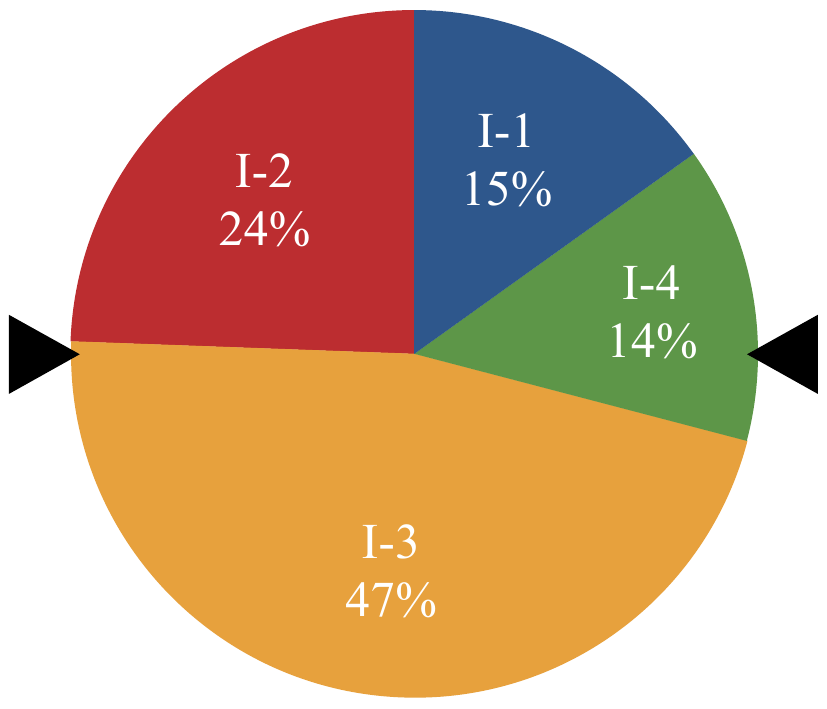} 
		{\color{gray} \hspace{1pt}\rule{0.4pt}{90pt}\hspace{1pt}}
	\includegraphics[width=0.235\linewidth]{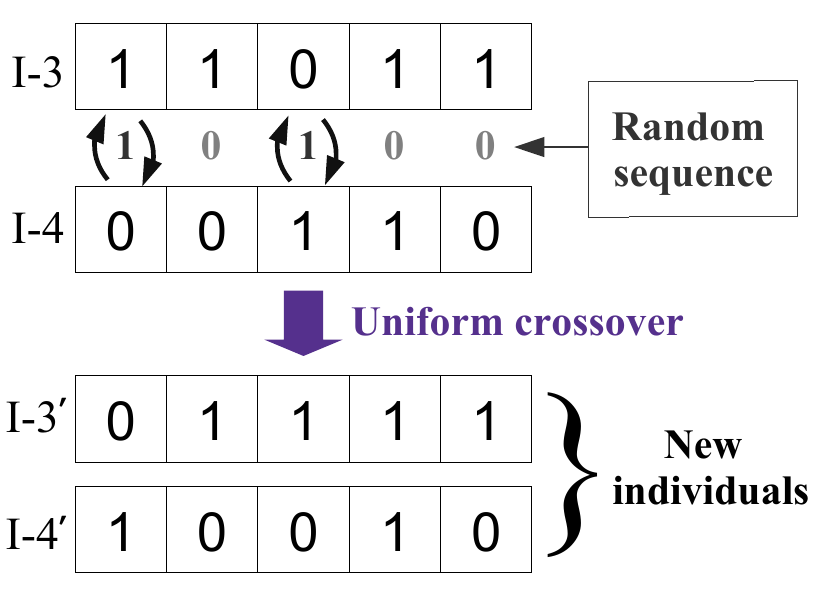}
		{\color{gray} \hspace{1pt}\rule{0.4pt}{90pt}\hspace{1pt}}
	\includegraphics[width=0.255\linewidth]{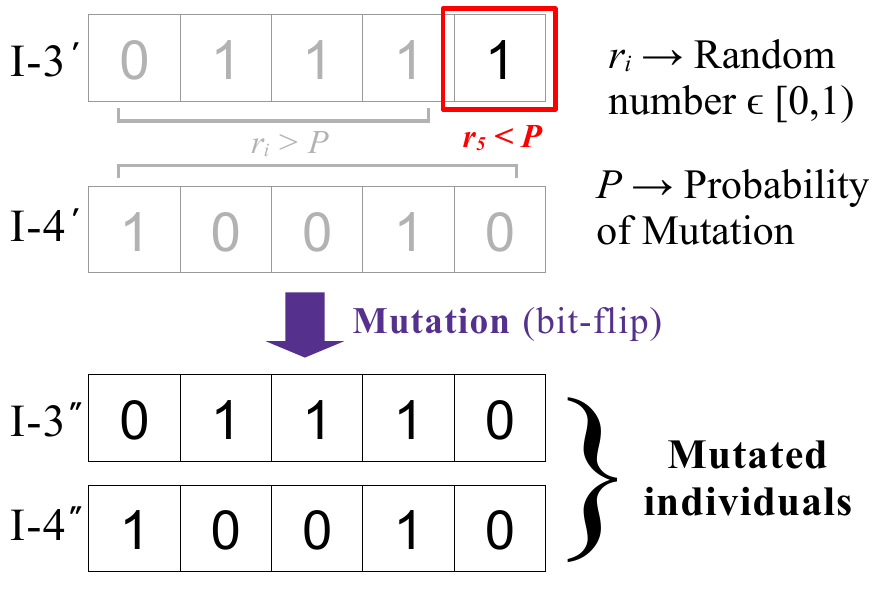}\\
	\hspace*{0.11\linewidth}(A) 
	\hspace*{0.22\linewidth}(B) 
	\hspace*{0.21\linewidth}(C)
	\hspace*{0.23\linewidth}(D)
	\hspace*{0.10\linewidth}
	\caption{A schematic representation of a toy-model of GA. (A) shows a population with 4 individuals and indicates the terminology within a GA framework. (B) shows a Roulette wheel with two fixed-points diametrically opposite pointing to I-3 and I-4. (C) shows a typical uniform crossover event on I-3 and I-4, producing the new individuals I-3$'$ and I-4$'$. These are the offspring which then get mutated into I-3$''$ and I-4$''$ as shown in (D). All the numbers including binary strings and probabilities are only meant for visualization of this toy-model.} \label{fig:GA-ovrw}
\end{figure*}

The goal of GA is to numerically find the global optimum of an objective function (also referred to as a ``cost function'') in the given domain of interest. GA works on the principle of convergence with variation. Diversity in a population is a necessary ingredient for the GA to succeed. Over the generations, individual solutions with higher and higher fitness dominate the population. Fig.~\ref{fig:GA-flow} shows the flowchart of a typical GA. In the following we briefly discuss the most common steps of GA. 

\begin{enumerate}
    \item \underline{Population initialization}: GA derives from the basic ideology of population genetics. Hence, GA is based on a ``population'' -- a set of possible solutions to the problem. According to the domain of search for the optimal solution, a population can be randomly created. The size of a population is a heuristic parameter.
    
    \item \underline{Fitness calculation}: A fitness function is chosen such that the fitter individuals will be closer to the optimum (in the functional space) than the less fit ones. The choice of a fitness function depends on the underlying objective and it is not unique. However, in order for a proper implementation of the algorithm, the fitness function should be positive and analytic, at least in the domain of interest.
    
    \item \underline{Parent selection}: This is one of the most crucial steps of the entire algorithm. As Natural selection dictates, the fitter individuals in the population are more likely to pass on their genes by mating than the less fit ones. To simulate this scenario, we must assign some weights to the individuals according to their fitness. One most common method is the fitness-proportionate assignment of weights (probabilities). In this step the pairs are selected for mating and producing the new generation. There is a set of methods for this. We will discuss some of them in section~\ref{sec:meth}.
    
    \item \underline{Crossover and Mutation}: From the parents selected in the previous step, we now make crossovers to generate new individuals, collectively called ``offspring''. The new individuals are occasionally mutated with a small amplitude. In our case, mutation refers to a discrepancy in the passage of genes from one generation to the next. Mutation is a crucial tool to bring in variety in the genetic material. We discuss the importance of mutation in section~\ref{sec:meth}. The offspring thus produced will constitute the next generation of our population.
    
    \item \underline{Termination}: Steps (ii)-(iv) are repeated in a loop for each new generation until the stopping criteria are reached/ GA has ``converged''. In general, a GA implementation is said to have converged if the fitness of the best individual(s) over the generations does not improve any further. The programmer can set a stop-point generation, hitting which will end the generation loop. This stop-point can be found by \textit{hit-and-trail}. Altogether, the fittest individual in the last generation is deemed the solution of the GA run.
\end{enumerate}

Fig.~\ref{fig:GA-ovrw} uses a toy-model population to visually understand the different terms and steps in our GA implementation. (We discuss it further in the due course.) In section~\ref{sec:meth}, we describe the methodology adopted for this implementation of GA-ILC with the details concerning the exact styles and the values of certain parameters.

%-------------------------
%-------------------------
\section{Methodology}
\label{sec:meth}
%-------------------------
%-------------------------

Python\footnote{Open-source Python programming language, see \url{https://www.python.org/}} is one of the most commonly used programming languages for scientific computation. It provides many open-source packages for various utilities. One of those packages is the python-port of HEALPix, HEALPy. It comes with many utilities we need for handling and analyzing the data. There are pixel utilities, spherical-harmonic transform utilities, visualization related functions and many miscellaneous tools for an effective manipulation and reduction of a HEALPix data-set. We have implemented the algorithm in Python 3 for numerical computations.

In this work, we implement GA to find the optimal weights to linearly combine the input maps such that the reduced variance of the clean map is minimized. The reduced variance is thus a cost function for our case-study of GA-ILC. This is a multivariable constrained optimization problem. Here, it is converted to an unconstrained optimization problem by a little manipulation of the method. The objective function to be optimized (minimized) is the (reduced) variance function (Eqn.~\eqref{eqn:sigma-sq-A}) 
\begin{eqnarray}\label{eqn:var-func}
	\sigma^2(\mathbf{w}) = \mathbf{w}^{\text{T}}\mathbf{Aw}.
\end{eqnarray}
The constraint is $\sum_{i=1}^{N}w = 1$. Alternatively, for the current work, one of the $N$ weights is found as $w_{i'} = 1 - \sum_{i \neq i'} w_i$. (It is the last weight of the spectrally-ordered system for our work.) Thus, the goal is to estimate $\mathbf{w}_{\text{GA}}$ that minimize $\sigma^2(\mathbf{w})$ using GA.

%-------------------------
\subsection{Implementation of GA-ILC}
\label{sub:implmn}
%-------------------------

\subsubsection{Terminology} \label{subsub:meth-term}
In Genetics, an individual can be represented by its genotypic as well as phenotypic characteristics. In our GA, the phenotype of an individual is a decimal number in the case of a single variable problem and a vector (array) of $n$ decimal numbers in the case of a multivariable problem. We chose the binary-string representation as our genotype. The required decimal accuracy is a fixed parameter initially chosen by the programmer. Depending upon the accuracy, each individual fraction can be converted to an integer by multiplying it with 10$^{\text{decimal-accuracy}}$ and rounding off. Then the decimal integer is converted into its binary representation. In the multivariable cases, each element of a vector is converted into a binary string and all of those strings are concatenated in the original order into a single large string. We call this string a ``chromosome''. The binary part of an individual element of this vector chromosome is called a ``segment'', each position of a chromosome is called a ``gene'', and the bit-value of each gene is called the ``allele'' of that gene. Fig.~\ref{fig:GA-ovrw} (A) depicts a schematic of this terminology. It shows an example one-variable population in the genotype representation. Performing the exact inverse operations on a chromosome will return the phenotype (decimal vector) of that individual.

\subsubsection{Population  Initialization} \label{subsub:meth-pop-init}
Now that the terminology is clear, we need to initialize a population. The size of the population, \texttt{Popsize}, is fixed by the programmer by \textit{hit-and-trial}. For our GA-ILC, we chose $[-1, 1]$ as the initial domain for each variable, hereinafter called ``priors''. (It must be noted that the priors can only restrict the initial population, or the first generation. As the generations progress, it is infeasible -- and quite unnecessary for GA-ILC -- to limit the individual solutions only to a small domain. The true limitation is inherently imposed by the largest decimal accuracy required for the problem.) The initial population is then uniformly randomly generated. The population at any stage of calculation is a 2D array with one dimension containing genes of a single chromosome and the other containing as many as \texttt{Popsize} chromosomes. The allele of each gene of each chromosome is determined by generating a (pseudo-) random number between 0 and 1, and according to whether it is less or greater than 1/2. This is, in some sense, the coin-toss determination and it is frequently used in this work. From the binary population, the decimal population is also found. At this point a check is made on whether all the variables in each individual fall within the priors imposed. The coin-toss initialization is carried out until all variables in all individuals lie within the priors. The number of variables in our GA-ILC, \texttt{N\_var}, is one less than the number of input maps since one of the weights is dependent on the rest of them.

\subsubsection{Fitness Calculation} \label{subsub:meth-fitness}
As discussed in subsection~\ref{sub:GA_form}, choice of a fitness function is not unique. Our goal in GA-ILC is to minimize the reduced variance, but fitness is to be maximized by definition. Hence, a suitable fitness function for the GA-ILC is chosen as
\begin{eqnarray}\label{eqn:fitness}
	\bar{f}(\mathbf{w}) = \frac{1}{\sigma^2(\mathbf{w})}.
\end{eqnarray}
Since the $\sigma^2$ is strictly positive, this fitness function satisfies the criteria of being a positive analytic function (in principle, the global minimum of $\sigma^2$ is the reduced variance of a pure CMB map).

\subsubsection{Parent Selection} \label{subsub:meth-parents}
As mentioned in subsection~\ref{sub:GA_form}, this step refers to the selection of pairs that will then undergo mating.  The set of individuals selected in this step will constitute a ``mating pool''. The size of the mating pool (in terms of individuals, not pairs) is the parameter \texttt{N\_pairs}. The value of this parameter can be fixed or it may vary over the generations (more on this in subsection~\ref{sub:diversity}).

\begin{figure}
	\begin{center}
		\includegraphics[width=0.9\linewidth]{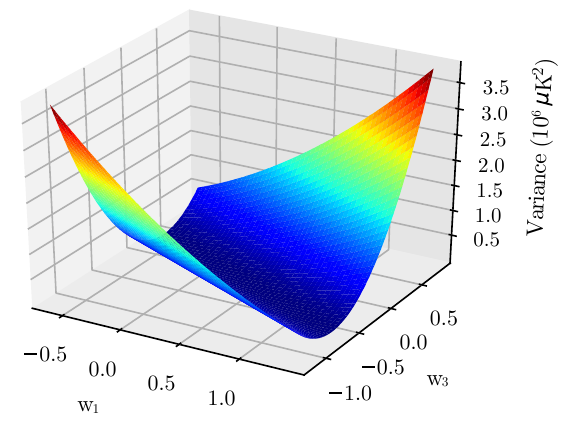}
	\end{center}
	\caption{The variance function in Eqn.~\eqref{eqn:var-func} plotted against two weights $w_1$ and $w_3$ while keeping the rest of the weights fixed to the SS weights. This is performed on the simulation of \texttt{seed} 100 low-resolution maps. A very broad valley can be seen close to the supposed global minimum.} \label{fig:w-vs-var}
\end{figure}
Fitness values indicate how better a solution is than another. According to natural selection, fitter individuals are more likely to pass on their genes to the next generation. This scenario is dubbed ``selection pressure''. To introduce selection pressure while selecting parents, some weight must be assigned to each individual in a population that will indicate how likely that individual is to be picked. To select pairs, i.e., two individuals at a time, a roulette wheel based approach is followed. The method is called ``Stochastic Universal Sampling'' or SUS in short. In SUS all individuals are allotted an (angular) area proportional to their weights on the roulette wheel. Since we need to select two individuals simultaneously, two fixed points (indicators) are there on the periphery of the wheel equiangularly spaced (opposite to each other in this case). A turn of the wheel is simulated by generating a pseudorandom number. Fig.~\ref{fig:GA-ovrw} (B) shows a roulette wheel in our toy-model population with some area given to all four individuals. Notice that the two fixed-points for picking pairs are diametrically opposite to each other. This way, an individual with low selection pressure may also get picked occasionally, maintaining the genetic diversity in GA.

In a fitness-proportionate (FP in short) selection of parents, the area assigned to each individual in a population on the wheel is directly proportional to their fitness. However, as the generations progress and the population approaches the desired optimum, the fitness value may get less and less diverse due to, say, a decreasing gradient of the objective function. In such cases, the selection pressure to the fitter individuals wanes since all the individuals will be assigned an \emph{almost} equal area on the wheel.

To circumvent this problem, another selection method such as ``Rank selection'' can be employed. In Rank selection, the individuals in a population are sorted in the decreasing order of their fitness values and they are assigned ranks (the fittest individual has the first rank). Then the areas on the roulette wheel are allotted according to their ranks, irrespective of their fitness values. For instance, in our GA-ILC while doing Rank selection, we allot area as the reciprocal of the rank (normalized; larger the rank, smaller the area). The selection is then done by the roulette wheel turnings under the SUS framework. This method brings the selection pressure back into the picture since the weight-gradient is preserved.

When the objective function in the domain of search is known to have inherently less gradient, Rank selection should be preferred from the start of the generation loop. If closer to the optimum point the function is less steep, then after a certain number of generations the selection type can be switched from FP to Rank selection. This switching-point generation is another GA parameter called \texttt{switch}. For example, the variance function in Eqn.~\ref{eqn:var-func} is plotted in Fig.~\ref{fig:w-vs-var} for a pair of weights while keeping the rest of the weights fixed to the SS-weights. It is done on one of the realizations in the Monte Carlo simulations we performed (we discuss it further in section~\ref{sec:MC-simul}. As can be seen in that figure, the variance function has a very broad valley close to the global minimum. We verified that the feature is present irrespective of the pair of weights chosen. This required us to switch between FP and rank selection types after a switch-point generation.

\subsubsection{Crossover and Mutation} \label{subsub:meth-cross-mut}
The crossover again comes with a variety of types. Some of the most popular types are one-point, two-point, $k$-point, and uniform crossovers. We chose uniform crossover as the universal type for our entire work. An instance of this type of crossover is schematically shown in Fig.~\ref{fig:GA-ovrw} (C) for our toy-model population. Under uniform crossover, the alleles of each gene are exchanged between the two parent chromosomes of a pair using the coin-toss determination technique. Thus, each pair of parents give rise to a pair of children. 

These children are mutated with a small amplitude, \texttt{P\_mut}. A mutation is introduced as the following. For each gene of each child chromosome, a pseudorandom number \texttt{r} in the range $[0, 1)$ is generated. If \texttt{r<P\_mut}, the allele of that gene is flipped (0$\to$1 and 1$\to$0). The parameter \texttt{P\_mut} is best chosen by the programmer by \textit{hit-and-trial}. Fig.~\ref{fig:GA-ovrw} (D) indicates one of the mutation incidents in the toy-model GA. After these steps, we have a set of offspring chromosomes at hand. The major purpose of introducing mutations is not to find fitter individuals, but to bring back genetic diversity which gets diminished over generations due to convergence. It must be noted that the mutation probability should not be very large, since it can lead the population away from convergence. Larger amplitude of mutation means that larger amount of `fit' individuals produced by crossover will be tempered with and brought away from the desired optimum (as only a very small fraction of mutated individuals contribute to an improvement in fitness over the generations). Besides, a large amount of mutation leads to the loss of the genetic inheritance (pass-on) feature. In our work, we explore the moderately low values of mutation probability ranging roughly around 0\% to 6.5\%. We address this further in section~\ref{sec:results}.

Depending upon the fitness of the children and the size of the population, a new generation is recruited by keeping \texttt{Popsize}-many fittest children into the population and discarding the rest of them. The newly deemed population will then undergo the same steps of parent selection and mating. Proper termination criteria in terms of the number of generations, \texttt{N\_gen}, is found by observing the convergence of GA-ILC and subject to the computational resources.

%--------------------
\subsection{Remarks on Genetic Diversity}
\label{sub:diversity}
%--------------------

It is particularly important to ensure a genetically diverse population at \emph{each} step of GA. The initial population is randomly initialized to have the maximum diversity possible. While applying selection pressure towards the fitter individuals, the less fit individuals should not be neglected entirely. A balanced reproduction is only possible if the entire population takes part in mating. We need to make sure that the least fit individuals also get to reproduce at least a couple times. A way to ensure this (in addition to using SUS method) is to make \texttt{N\_pairs} $\sim  \lceil1 /(\text{min}(\{p_i\}))\rceil$ where $\{p_i\}$ is the set of probabilities (weights) assigned to the individuals and $\lceil\rceil$ is the ceiling function. If this is less than half \texttt{Popsize}, \texttt{N\_pairs} can simply be \texttt{Popsize}. Due to the limited resources and increasing requirement of computing power, the bottleneck must be applied to filter-out the fitter individuals after reproduction and preserve the size of the population at each generation. This results in only a \emph{slightly} reduced genetic diversity, which is harmless considering the variety of children produced by mutation. Mutation, thus, is a very crucial ingredient in our GA. Indeed, a mild mutation prevents GA from premature convergence. A GA-ILC implementation without any mutation loses the genetic diversity almost completely as it moves towards the global optimum and causes it to converge to some false local optimum which is not actually present. It is a partly computational -- partly fundamental defect (in that a non-diverse, self-breeding population eventually ceases to survive under natural selection). In Population Genetics, occasional mutation may, by pure chance, introduce a better trait for survival and is preferred through generations. We studied this very interesting ingredient in more detail in our work. We discuss this further in sections \ref{sec:MC-simul} and \ref{sec:results}.

%-------------------------
%-------------------------
\section{Validation of the GA implementation}
\label{sec:valid}
%-------------------------
%-------------------------

%
\begin{figure}
	\begin{center}
		\includegraphics[width=\linewidth]{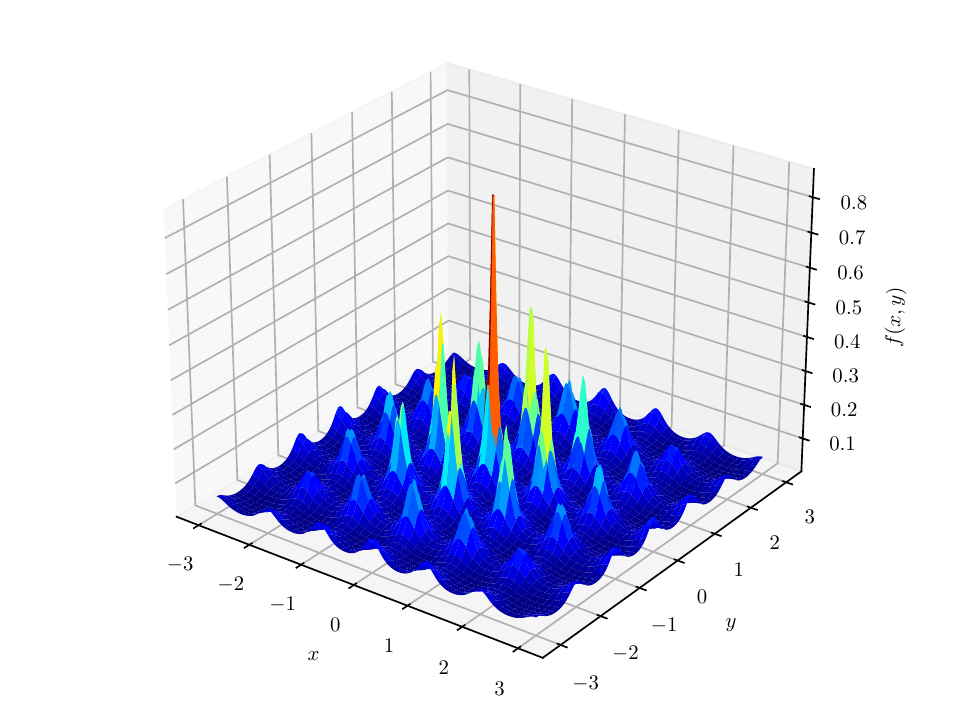}
	\end{center}
	\caption{The two-variable case of the trial function in Eqn.~\eqref{eqn:trial-fn} within an appropriate domain of interest. Observe the highly multimodal nature with multiple local maxima located closely together.}\label{fig:trial-fn}
\end{figure}

Implementation of GA is a complicated task. In particular, the programmer needs to find the optimal values of various GA parameters mostly by \textit{hit-and-trial} as mentioned quite frequently throughout subsection~\ref{sub:implmn}. The most common GA parameters are the population size \texttt{Popsize}, number of total generations \texttt{N\_gen}, decimal accuracy required \texttt{Dec}, probability of mutation \texttt{P\_mut}, etc. Hence, it is wise to start with a very simple implementation and then move towards more complicated problems. We have employed this strategy for this work. We initially implemented GA for single variable test-problems. Then we moved on to multivariable test-problems and we obtained satisfactory results in both of them. The test function we chose is
\begin{eqnarray}\label{eqn:trial-fn}
	f(\mathbf{x}) = \frac{1}{1 + \sum_{i=1}^{N_{\text{var}}}[ x_i^2 - 10(\cos(2\pi x_i) - 1)]},
\end{eqnarray}
with the same function as the fitness function. This function is well-defined for any number of variables. The two-variable case of this function is shown in fig.~\ref{fig:trial-fn} for visualization. It is observed from the figure that this trial function is a highly multimodal function \textit{i.e.} it has multiple local maxima with a single global maximum at $\mathbf{x = 0}$ with $f(\mathbf{0}) = 1$, within the domain of interest as plotted therein. Also, it is a steep function with large gradient. Hence, for such implementations, no mutation was necessary (in other words, \texttt{P\_mut} = 0) and after $\sim$60\% of total generations, the selection type was switched from FP to rank selection to compensate for having no mutation. The best of those results for different $N_{\text{var}}$ are summarized in table~\ref{tab:primary-GA}. Table \ref{tab:primary-GA-11} shows the results obtained for the fixed number of variables in the trial function -- which is the same as that in our GA-ILC implementation -- with different population sizes. As expected, the best function value improves as we increase the size of the population. The results of the trials as discussed here are sufficient to conclude that our GA implementation is robust and accurate in the multivariable scenario of our GA-ILC method, and that it is safe and desirable to incorporate ILC with it in the manner described in section~\ref{sec:meth}.
\begin{table}
	\caption{Results of the trial implementation for different number of variables. Notice that, for larger and larger number of variables, to get the best results, we need to have a larger and larger population.}\label{tab:primary-GA}
	\centering 
	\begin{tabular}{|c|c|c|c|}
		\hline
		$N_{\text{var}}$ & \texttt{Popsize} & \texttt{N\_gen} & Best $f(\mathbf{x^*})$\\
		\hline
		1 	& 200 	& 60	&1.0 \\
		2	& 200	& 60 	&1.0\\
		5	& 400	& 200 	&1.0\\
		10	& 600	& 250	&1.0\\
		11	& 600	& 250	&1.0\\
		12	& 600 	& 250	&1.0\\
		\hline
	\end{tabular}
	
	\caption{Variation of the final function value (in the trial implementation) \emph{w.r.t} \texttt{Popsize} for fixed number of generations \texttt{N\_gen} = 250 and $N_{\text{var}}$ = 11.
		Notice the very bad function value for small \texttt{Popsize}.}\label{tab:primary-GA-11}
	\centering
	\begin{tabular}{|c|c|}
		\hline
		\texttt{Popsize}  	& Best $f(\mathbf{x^*})$\\
		\hline 
		400		& 0.5011925167207855\\
		450		& 1.0000000000000000\\
		500		& 0.9999999999980158\\
		550		& 0.9999999999682601\\
		600		& 1.0000000000000000\\
		650		& 1.0000000000000000\\
		\hline
	\end{tabular}
\end{table}
\begin{figure*}%[!ht]
	\begin{center}
		\includegraphics[width=0.48\linewidth]{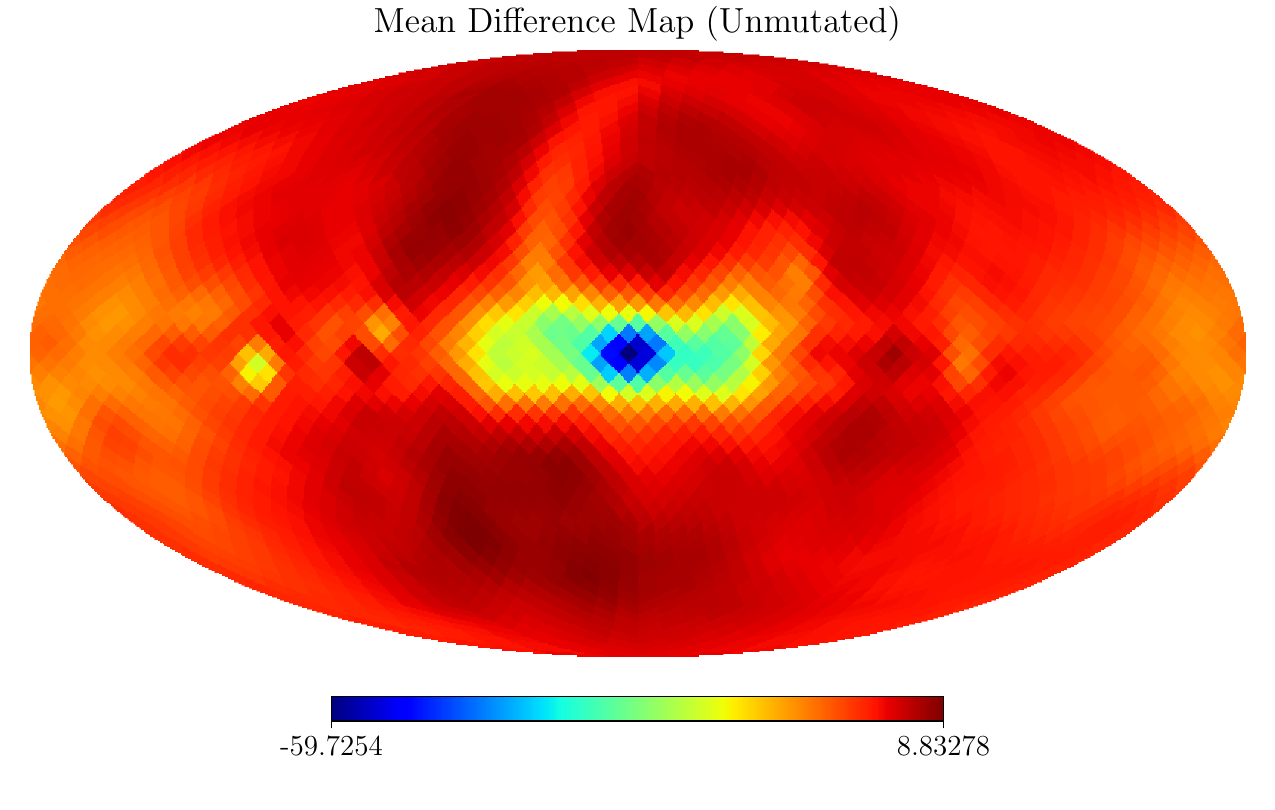}
		\includegraphics[width=0.48\linewidth]{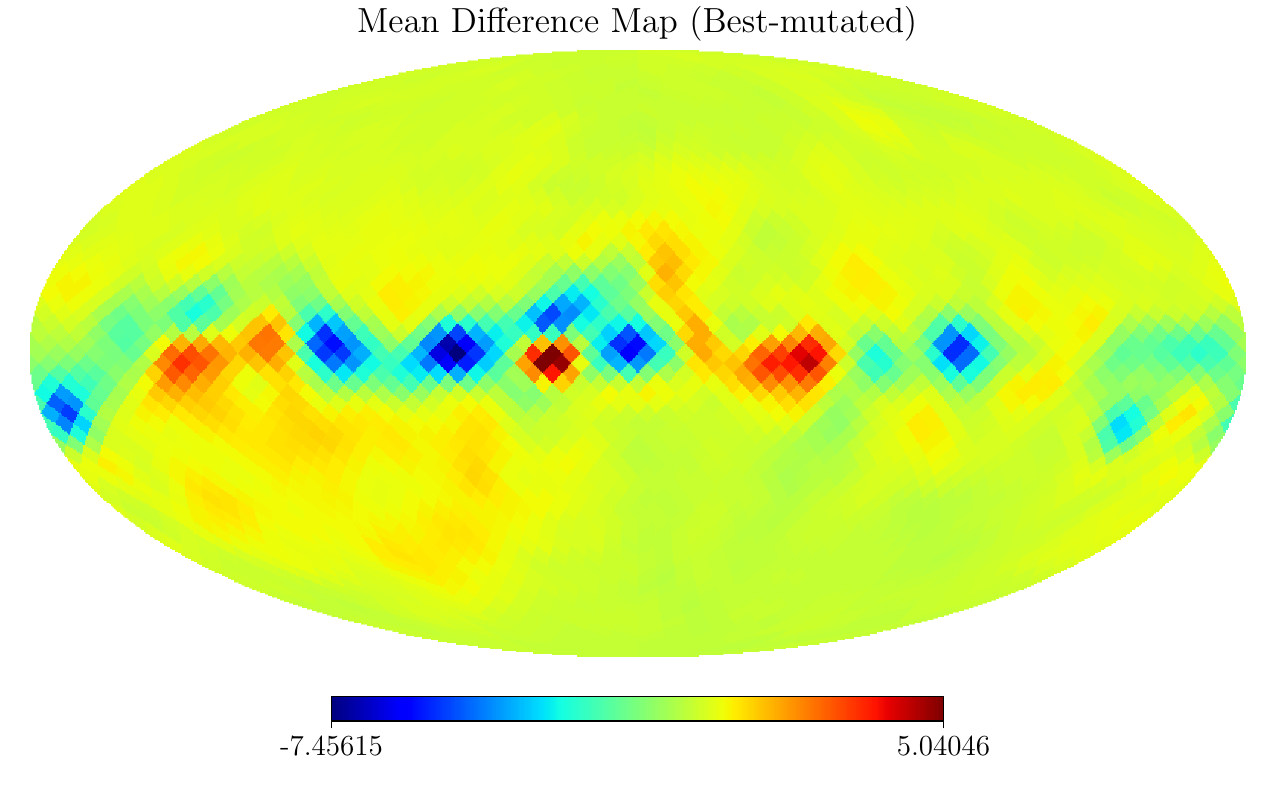}
		\includegraphics[width=0.48\linewidth]{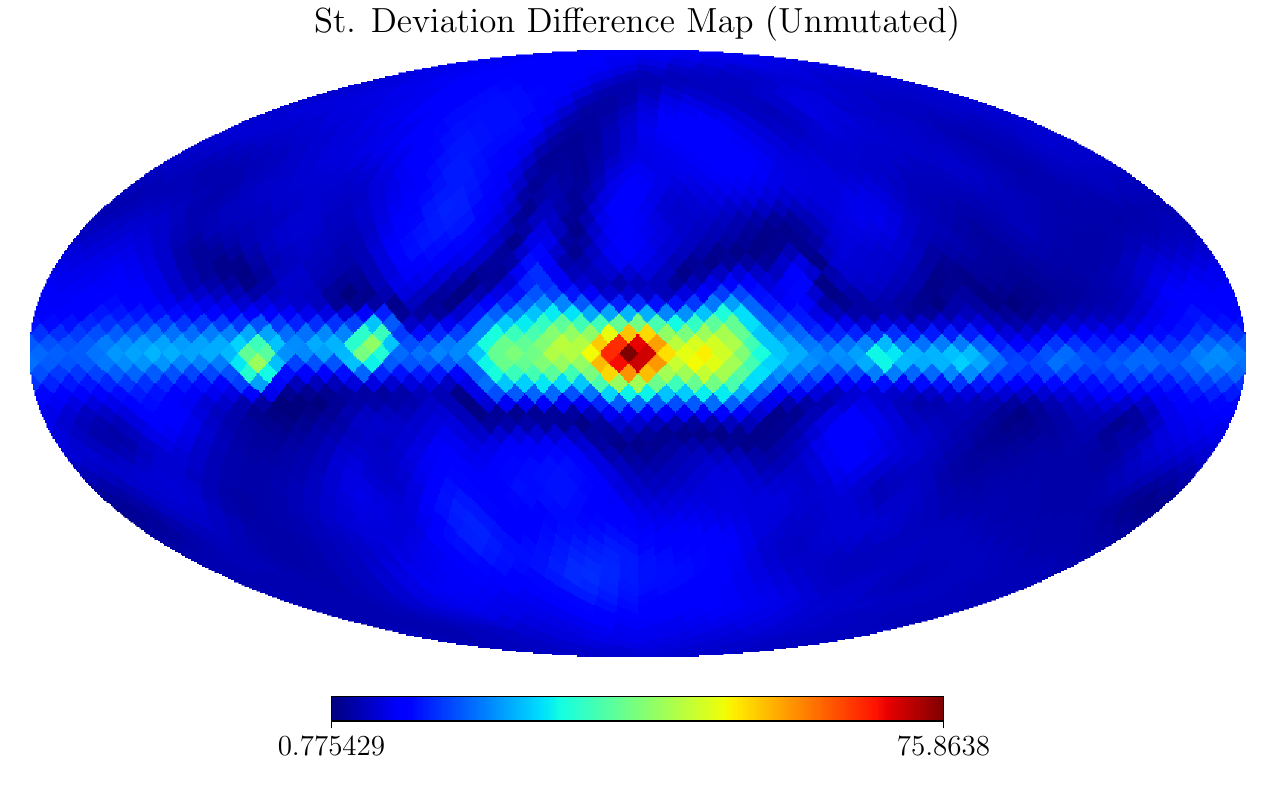}
		\includegraphics[width=0.48\linewidth]{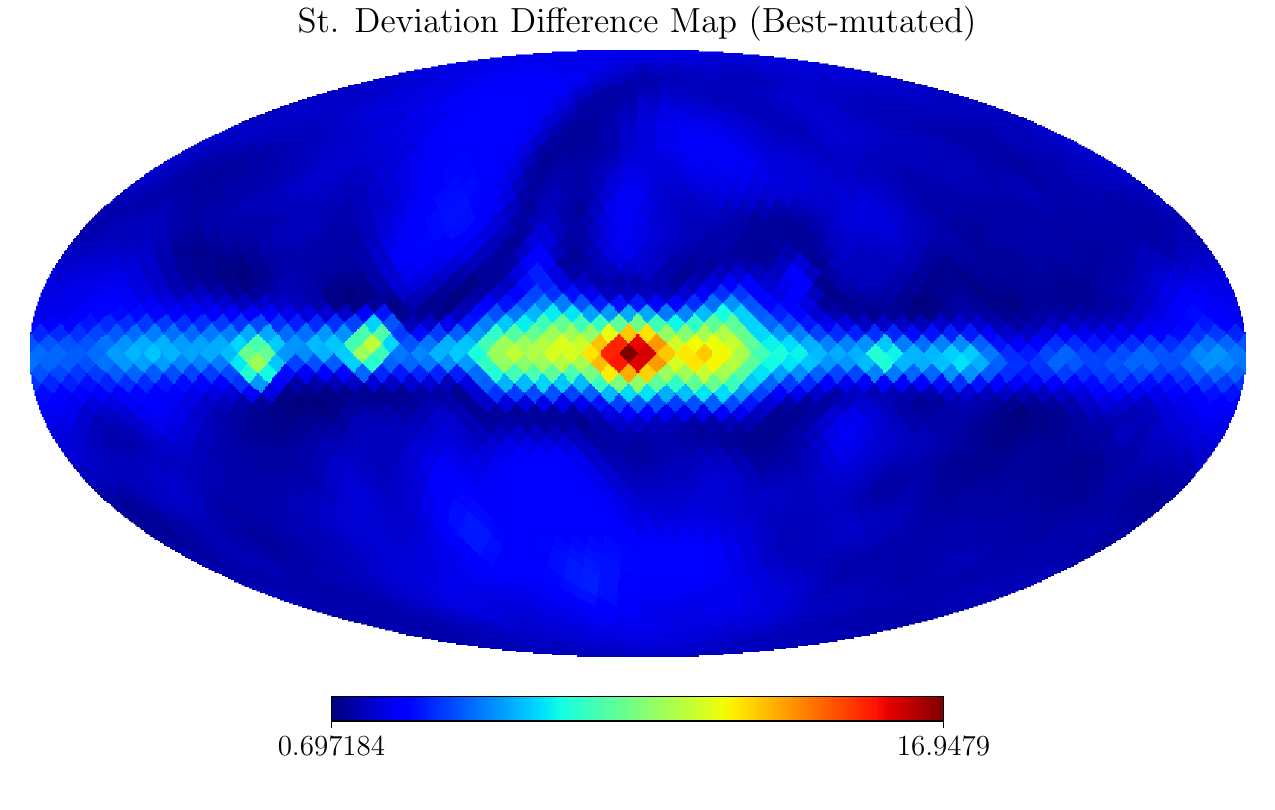}
	\end{center}
	\caption{The statistical difference maps from the 200 Monte Carlo simulations. The top panel shows the mean difference maps and the bottom panel shows the standard deviation difference maps. The maps on the left side show the statistics in the case of no mutation, and large foreground residuals can be seen close to the galactic equator. The maps on the right side represent the statistics of all the best (optimally) mutated cases. Only very small residual contamination seems to be present around the galactic equator therein. The unit of the values is $\mu$K.}
	\label{fig:simul-diff-maps}
\end{figure*}
\begin{figure}%[!ht]
	\includegraphics[width=\columnwidth]{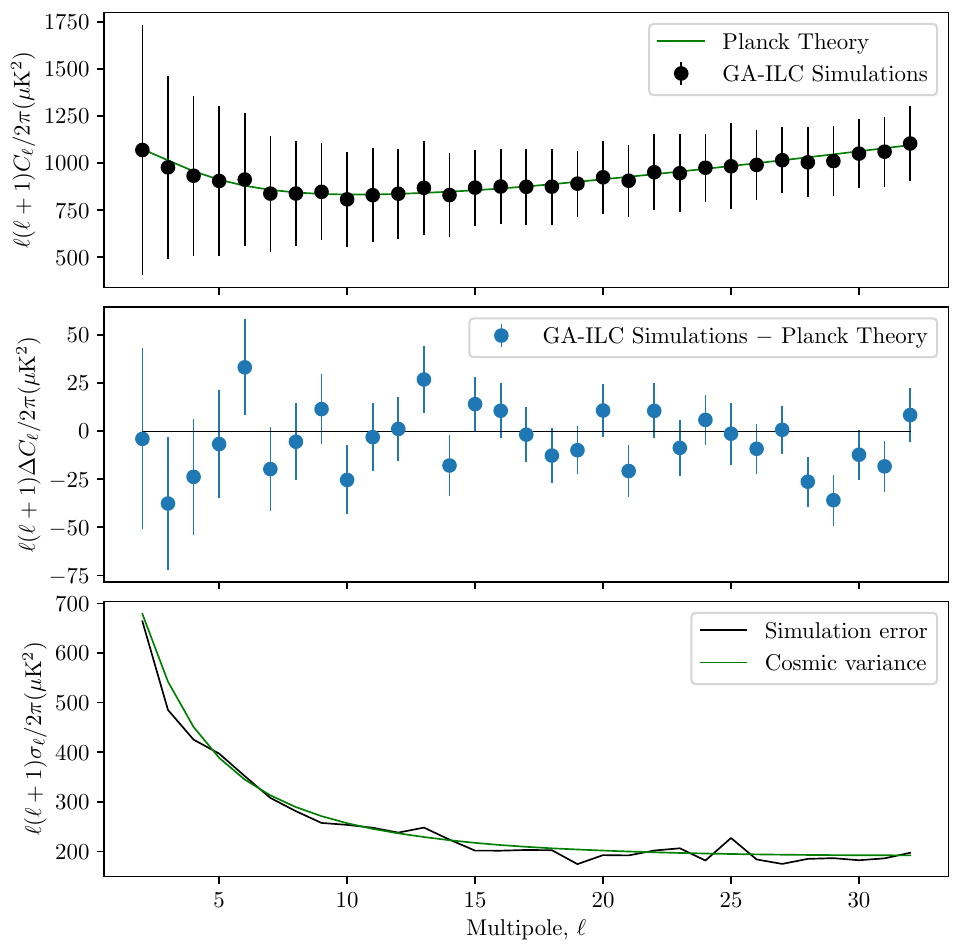}
	\caption{The angular power spectrum of the GA-ILC clean simulated mean difference map overplotted with the Planck theoretical power spectrum. Note that the beam and pixel effects are removed. Middle: The difference of the power spectra of GA simulations and Planck theory with errors in mean-estimation. It can be seen that the deviation is within 3$\sigma_{\mu}$ level which indicates that there is no bias. Bottom: The standard deviation errors in the angular power spectrum of the final mean difference map. Notice that these errors conform with the minimal errors induced by the cosmic variance, $\Delta C_{\ell, \text{cv}} = \sqrt{\frac{2}{2\ell+1}}C_{\ell}$.}
	\label{fig:simul-diff-cls}
\end{figure}
%%

%-------------------------
%-------------------------
\section{Monte Carlo Simulations}
\label{sec:MC-simul}
%-------------------------
%-------------------------	

As a statistical test for our GA-ILC method, we performed Monte Carlo simulations using WMAP and Planck simulated maps. We followed the methodology of \cite{Sudevan_2018} \textsection 7.1 to generate the input foreground and CMB maps for the GA-ILC. We chose a sample of size 200 for the simulations, wherein each individual -- denoted by a parameter called \texttt{seed}
\footnote{This is the same \texttt{seed} parameter used to simulate the input CMB map from the HEALPix \texttt{synfast} routine.} 
-- consists of a \textit{distinct} set of 12 multifrequency simulated maps (foreground + CMB). Of the 12 input maps, the 5 WMAP frequency bands are K (23 GHz), K$_{\text{a}}$ (33 GHz), Q (41 GHz), V (61 GHz), W (94 GHz); and the seven Planck frequency bands are 30, 44, 70, 100, 143, 217 and 353 GHz. These simulated maps are 9\textdegree  beam-smoothed. Some of the GA parameters were kept fixed for the simulations, namely, \texttt{Popsize} = 400, \texttt{N\_gen} = 400, decimal accuracy \texttt{Dec} = 7, and switching between FP and rank selection types after \texttt{switch} generation \# 20, all of them optimally chosen. 

%%%%
\begin{figure}
	\includegraphics[width=\columnwidth]{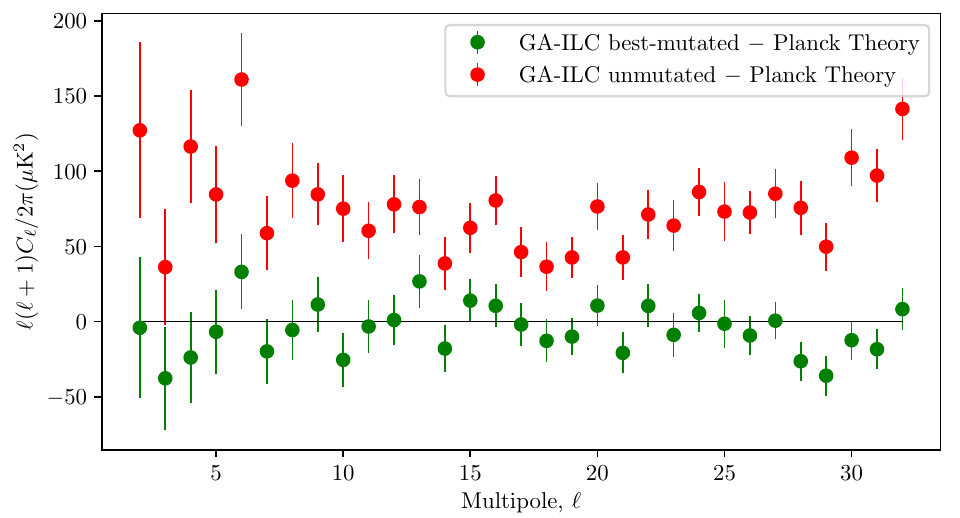}
	\caption{Comparison of the differential angular power spectra of the mean difference maps of GA-ILC without mutation and with best mutations. The plot with mutation included is the same as that in the middle panel in fig.~\ref{fig:simul-diff-cls}. We can observe that the unmutated case has significantly higher power at more-or-less all multipoles (with the deviation $\gtrsim 3\sigma_{\mu}$ at most multipoles) studied in this work.}
	\label{fig:cl_sim_mut-v-unmut}
\end{figure}
%%%%

To explore the effect of mutation on the results and also to find an optimal amount of mutation for an implementation, GA-ILC was run for 128 different values of \texttt{P\_mut} starting from 0\% (no mutation), increasing linearly in steps of 0.05\%, for each \texttt{seed} value. Note that each individual GA-ILC with unique values of mutation and \texttt{seed} is called an ``instance'' of GA-ILC. This was achieved by parallelizing all the 128 individual GA-ILC instances of a particular \texttt{seed} with different mutation values on a total of 128 cores of a HPC cluster\footnote{HPC cluster ``Kanad'' of IISER Bhopal, \url{http://atlas.iiserb.ac.in/index.html}}. The 200 simulations were run in series. The average time of execution of a GA-ILC instance thence came out to be 1261.70 seconds, or roughly 21 minutes. The difference between the clean map and the input CMB map for an instance is simply called a difference map here. First we found the statistical difference maps of all the 200 \emph{unmutated} (non-optimal) cases and then all the \emph{best-mutated} (optimal) cases. (In the optimal scenario, for each of the 200 \texttt{seeds}, the instance of \texttt{P\_mut} with the minimum final reduced variance is deemed the ``best-mutated''.) Fig.~\ref{fig:simul-diff-maps} shows the statistical (mean and standard deviation) difference maps in both, the unmutated and the best-mutated cases. The unmutated (non-optimal) case evidently carries large residual foregrounds close to the galactic plane. The best-mutated case also carries residual contamination in the similar areas, however, it has a very low amplitude and it is limited to small regions. 

%%%%
\begin{figure}
	\includegraphics[width=\columnwidth]{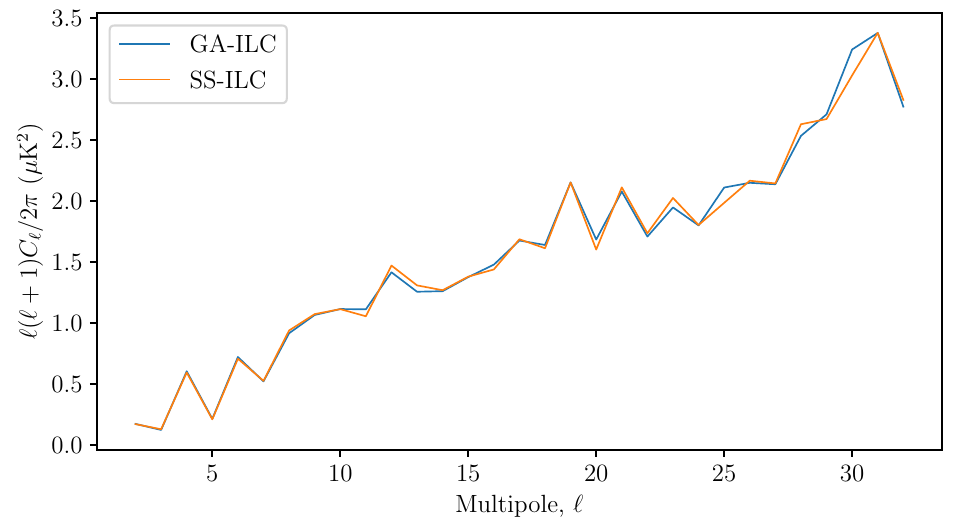}
	\caption{Angular power spectra of the mean difference maps of GA-ILC and SS-ILC from their respective simulated analyses for comparison. The absolute residual foregrounds are quite small and can be seen to agree excellently with each other at all multipoles studied in this work.}
	\label{fig:residual-cl-comp}
\end{figure}
%%%%

The angular power spectrum corresponding to the simulated mean clean map along with the standard deviation errors is overplotted with the Planck theoretical CMB power spectrum in the top panel of fig.~\ref{fig:simul-diff-cls}. In the middle panel of the figure, the mean difference power spectrum is plotted with the errors in the estimation of mean ($\sigma_{\mu, \ell}$). We observe that the estimated power spectrum is within a 3$\sigma_{\mu}$ deviation to the theoretical CMB power spectrum. This confirms that there is no bias in the entire multipole range $2<\ell<32$ under study. In the bottom panel of the same figure, the standard deviation errors ($\sigma_{\ell}$) are overplotted with the cosmic-variance induced errors. The observed errors also agree well with the expected minimal errors purely of the cosmic origin.

\begin{figure}
	\includegraphics[width=0.99\columnwidth]{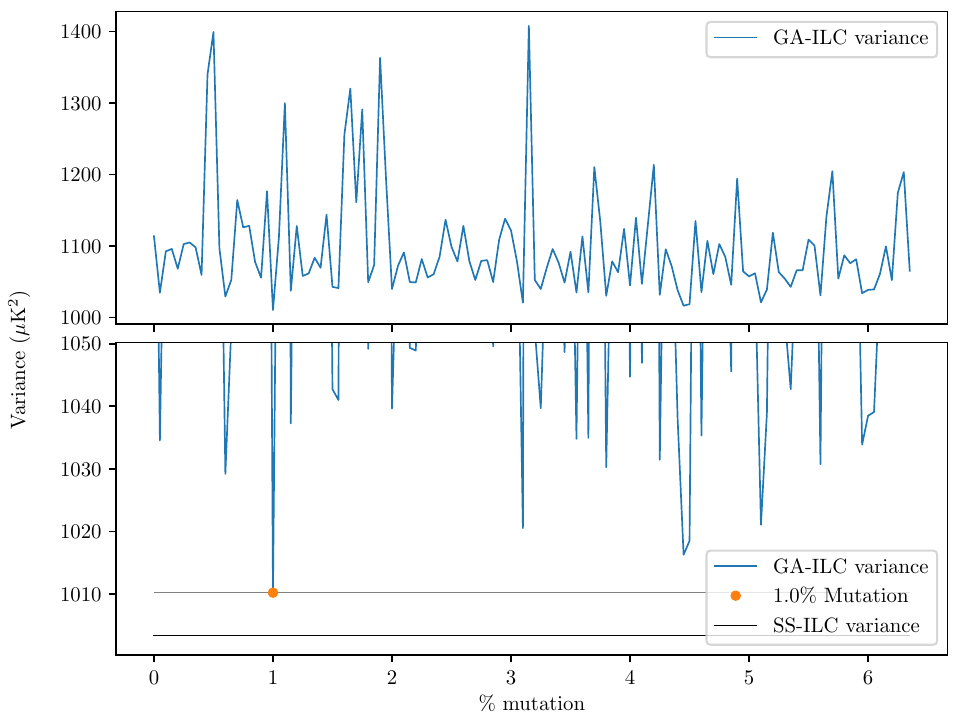}
	\caption{The minimal variance obtained by GA-ILC on data against the coefficient of mutation. The minimum of these minimal values -- that occurs at a low mutation by 1\% -- is also shown distinctively. Bottom: a zoomed-in version for a closer look at the variation. The variance derived from SS-ILC is shown for reference.} \label{fig:GA18.4_mut-plot}
\end{figure}
\begin{figure}
	\centering
	\includegraphics[width=\linewidth]{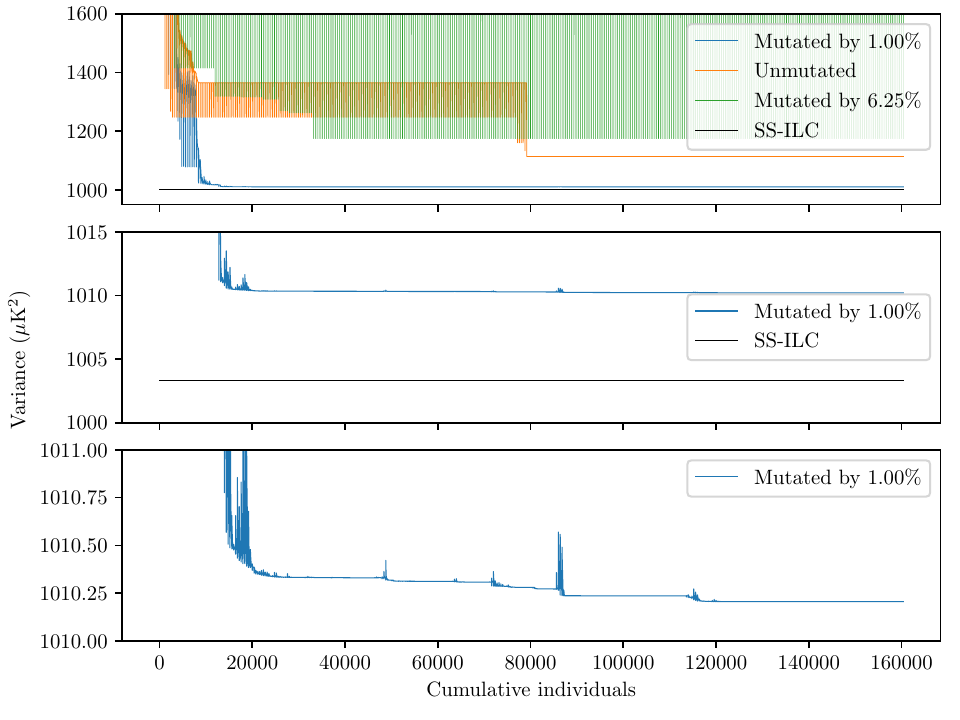}
	\caption{Top: the overplot of the traces of unmutated, optimally mutated and non-optimally mutated cases. The reduced variance of the clean map derived using SS-ILC is plotted as a reference value. The optimally mutated case is the best among all the 128 mutation values with \texttt{P\_mut} = 1\% implemented on data. As an example, a non-optimal case of mutation with \texttt{P\_mut} = 6.25\% is also shown. Middle and Bottom: closer look at the convergence in the mutated case. Note that the populations plotted here are sorted in ascending order of fitness. In the bottom panel we can observe the ``episodes'' of natural selection in our artificial population, during each of which the variance decreases (improves) over the course of a few generations.} \label{fig:GA18.4_trace-plot}	
\end{figure}

We illustrate the importance (and necessity) of mutation with fig.~\ref{fig:cl_sim_mut-v-unmut}. We plot the difference of angular power spectra between our simulations (GA-ILC) and the Planck theory. The two cases in the plot are distinguished by the presence and absence of mutation. The best-mutated case is compiled with the best cases for all 200 seeds (the same as the middle panel of fig.~\ref{fig:simul-diff-cls}). When there is no mutation included, the power is apparently quite high at all multipoles, with the deviation $\gtrsim 3\sigma_{\mu}$ at most multipoles. On the other hand, the best-mutated case of power spectra seems to agree quite well with the theoretical $C_{\ell}$. Indeed, mutation is a key ingredient in preventing premature convergence and guiding the population towards the global optimum.

To further investigate the residual foreground in the clean maps in the simulations, we compare the angular power spectra of the mean difference maps of GA-ILC with those of SS-ILC (from their respective MC simulations). As seen in fig.~\ref{fig:residual-cl-comp}, the residual foregrounds are small, not exceeding 3.5 $\mu$K$^2$, at all multipoles and they also agree with each other up to a very desirable degree.

%-------------------------
%-------------------------
\section{Application on WMAP and Planck data}
\label{sec:results}
%-------------------------
%-------------------------

We applied our GA-ILC on 12 multifrequency low-resolution input maps observed by WMAP and Planck to find an optimal clean map as output (we refer to this as the implementation on ``data'' as shorthand; the 12 input frequencies are those mentioned in section~\ref{sec:MC-simul}). GA parameters like \texttt{Popsize}, \texttt{N\_gen}, \texttt{Dec}, and \texttt{switch} were kept at the fixed values, same as the Monte Carlo simulations. Similarly to the simulations, we applied GA-ILC with 128 values of \texttt{P\_mut}, from 0\% to 6.35\% in steps of 0.05\%, to analyze the final reduced variance in each case. The best case of all these 128 was deemed to represent the best version of our GA-ILC. A moderately small value of 1.0\% is the best mutation value in this case. The plot of final reduced variance against the mutation probability is shown in fig.~\ref{fig:GA18.4_mut-plot}. The figure indicates that there exists no analytical relationship between the reduced variance and the amount of mutation. Indeed, the stochastic nature of the algorithm is clearly evident here. 

\begin{table}
	\caption{Comparison of the analytical SS-ILC weights and GA-ILC weights for the data-implementation.}\label{tab:weights-SS-GA}
	\centering 
	\begin{tabular}{|c|r|r|}
		\hline
		Frequency $i$ (GHz) & $w_{i, \text{SS}}$ & $w_{i, \text{GA}}$\\
		\hline
		23 	& -0.0934664   & -0.0104064	   \\
		30	&  0.2222572	& -0.0521856     \\
		33	&  0.4325201	&  0.6706560 	  \\
		41	& -0.3916232	& -0.4068993	  \\
		44	& -0.8638316  &   -0.7575169	   \\
		61	& -0.1091493 	&   0.1048943	     \\
		70  &  0.1897659   &  -0.9213568   \\
		94  &  0.4054625  &   0.3993862    \\
		100 &  0.8878745   &   1.3724032     \\
		143 &  0.9024152   &   1.5165824      \\
		217 & -0.6072505  &  -0.9606785      \\
		353 &  0.0250219  &   0.0451214        \\
		\hline
	\end{tabular}
\end{table}

\begin{figure}
	\includegraphics[width=0.99\columnwidth]{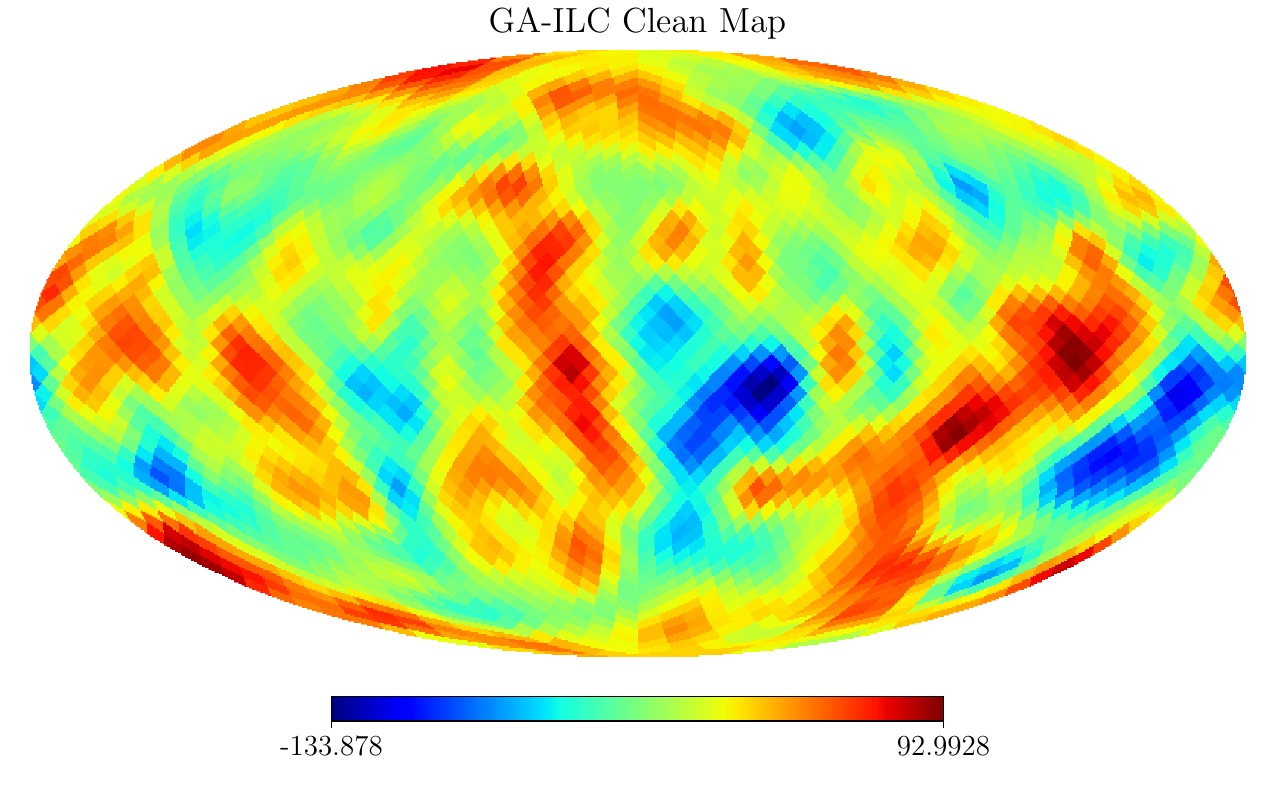}
	\includegraphics[width=0.99\columnwidth]{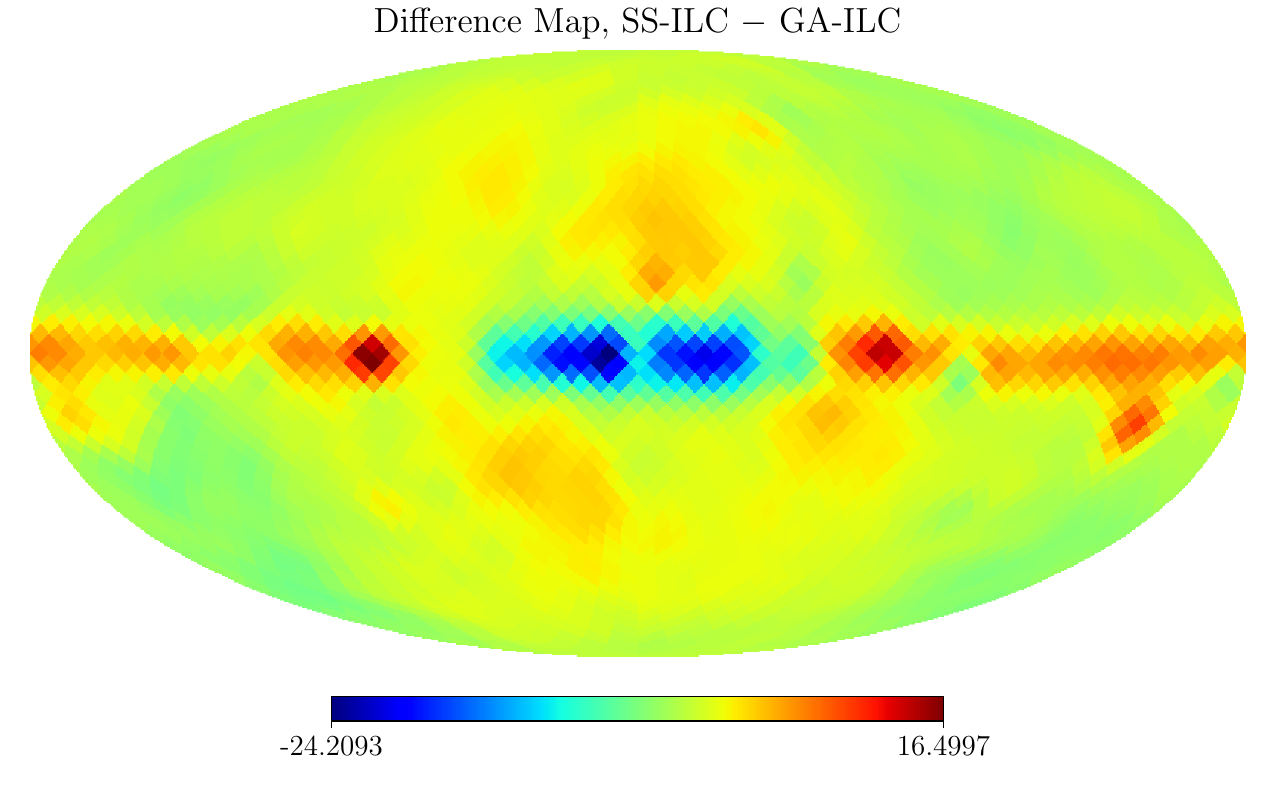}
	\caption{The clean map produced by the GA-ILC on WMAP and Planck data with 1\% mutation. Bottom: Difference map of GA-ILC and the SS-ILC. All the values are in $\mu$K. Notice that the residual contamination is very small and exists only close to the galactic plane. In other parts of the sky, a very good agreement can be seen.} \label{fig:GA18.4_clean-maps}
\end{figure}

\begin{figure}
	\includegraphics[width=\linewidth]{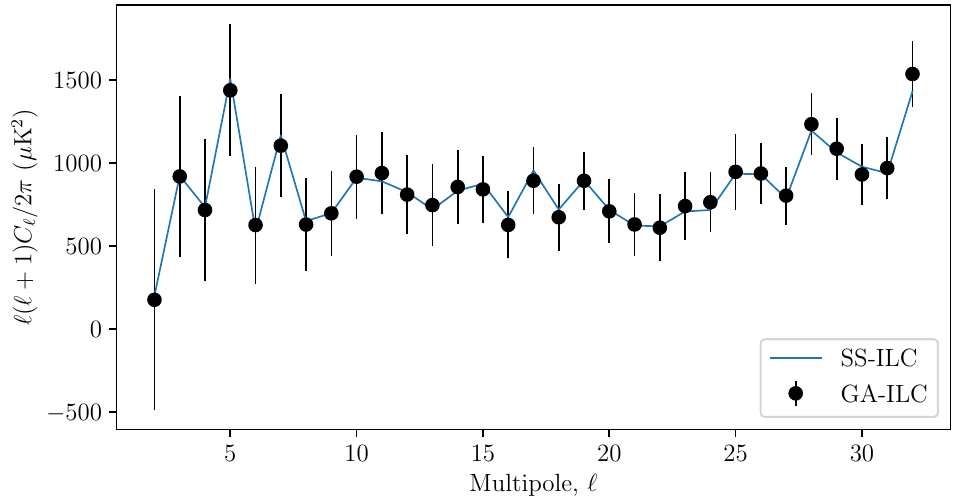}
	\caption{The clean angular power spectrum produced by GA-ILC implemented on data with 1\% mutation. The reconstruction errors in GA-ILC here are found using 200 Monte Carlo simulations. The two power spectra seem to agree quite well with each other.} \label{fig:GA18.4_clean-cls}
\end{figure}

We plotted the variance values against the cumulative individuals through generations (the population is sorted in the ascending order of fitness). This type of plots are called ``trace-plots'' here. These plots help us understand the convergence of GA and the effects of various parameters on it. Fig.~\ref{fig:GA18.4_trace-plot} shows an example of trace-plots. The top panel of this figure contains three distinct cases of mutation: (i) the unmutated case, (ii) the optimally mutated case with \texttt{P\_mut} = 1.0\%, and (iii) a non-optimally mutated case with \texttt{P\_mut} = 6.25\%. Therein we can see that the population without any mutation converges prematurely due to the loss of genetic diversity as there exists a broad valley close to the global optimum. On the other hand, the optimally mutated case is able to find better solutions since it is able to explore the variable space. It is also evident that even when the unmutated case has long since converged, the mutated case is slowly able to find better and better solutions after generations. The non-optimally mutated case is observed to introduce a very large variance in the fitness value, and therefore has the tendency to hinder the whole population from reaching the global optimal solutions. Indeed a balance between genetic diversity and genetic inheritance feature is necessary to be kept. 
	
In the bottom panel of fig.~\ref{fig:GA18.4_trace-plot}, the reduced variance seems to be decreasing slowly in irregularly spaced steps, e.g., small drops in reduced variance can be observed around 50000 and 70000 cumulative individuals. These steps are actually sets of many consecutive generations. At the beginning of each of these sets, mutation is able to find some slightly fitter solution(s) -- the fittest so far -- owing to a sudden increase (within a desirable bracket) in genetic diversity. Over the course of many upcoming generations within that set, the newly found fittest individual is favored for reproduction and so the whole population advances in the evolutionary sense. This kind of `natural selection episodes' are common in biological evolution which might lead to bifurcations and origination of new species. Hence, it is fascinating to observe such episodes even in our synthetic evolutionary settings. 

Fig.~\ref{fig:GA18.4_clean-maps} shows the clean CMB map produced by GA-ILC on data. Therein the bottom panel shows the difference map of GA-ILC and SS-ILC \citep{Sudevan_2018}. It confirms that the minimal residual contamination is present in the GA-ILC clean map and that it tends to occupy small areas close to the galactic plane. The clean angular power spectrum calculated from the GA-ILC clean map (data implementation) is overplotted with SS-ILC clean power spectrum in fig.~\ref{fig:GA18.4_clean-cls}. We observe that the two match very well with each other. This also indicates that the purely numerical GA-ILC technique gives as good results as SS-ILC with analytical expression of weights.

\begin{figure*}
	\includegraphics[width=0.49\linewidth]{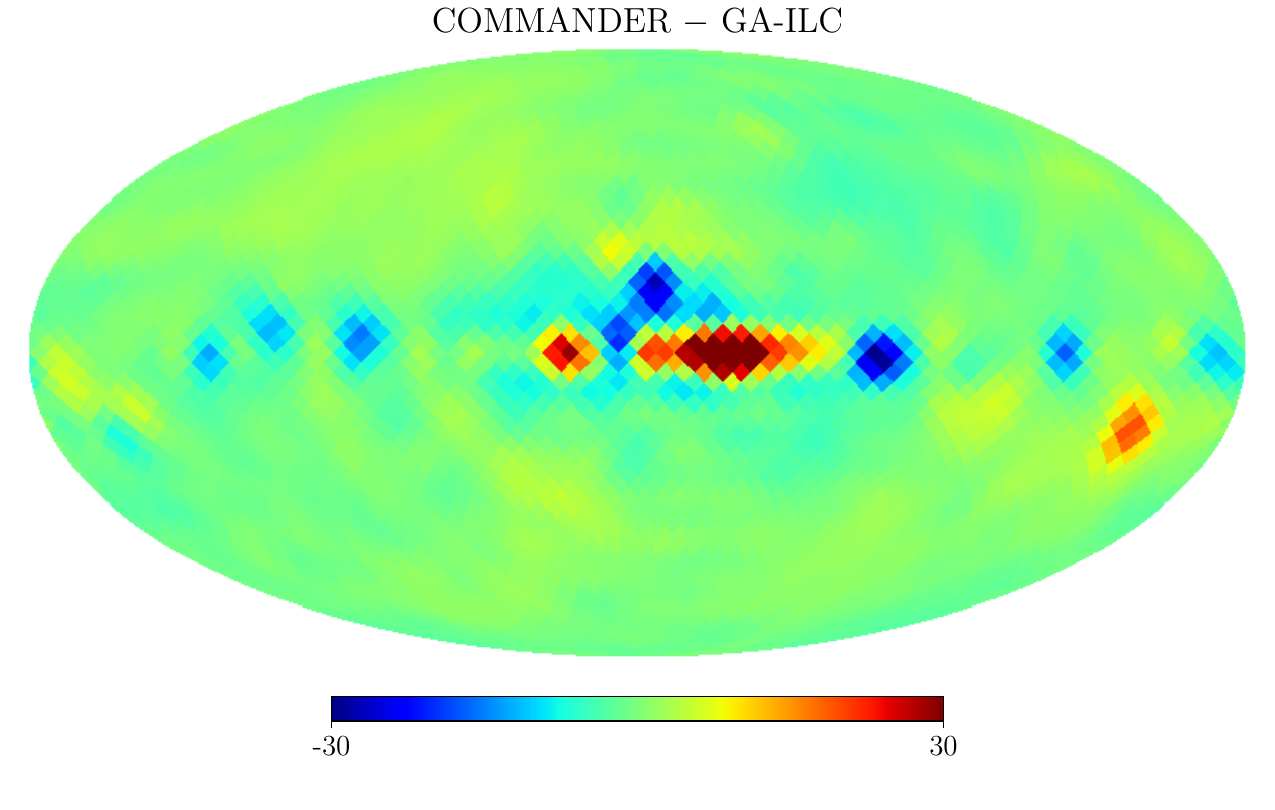}
	\includegraphics[width=0.49\linewidth]{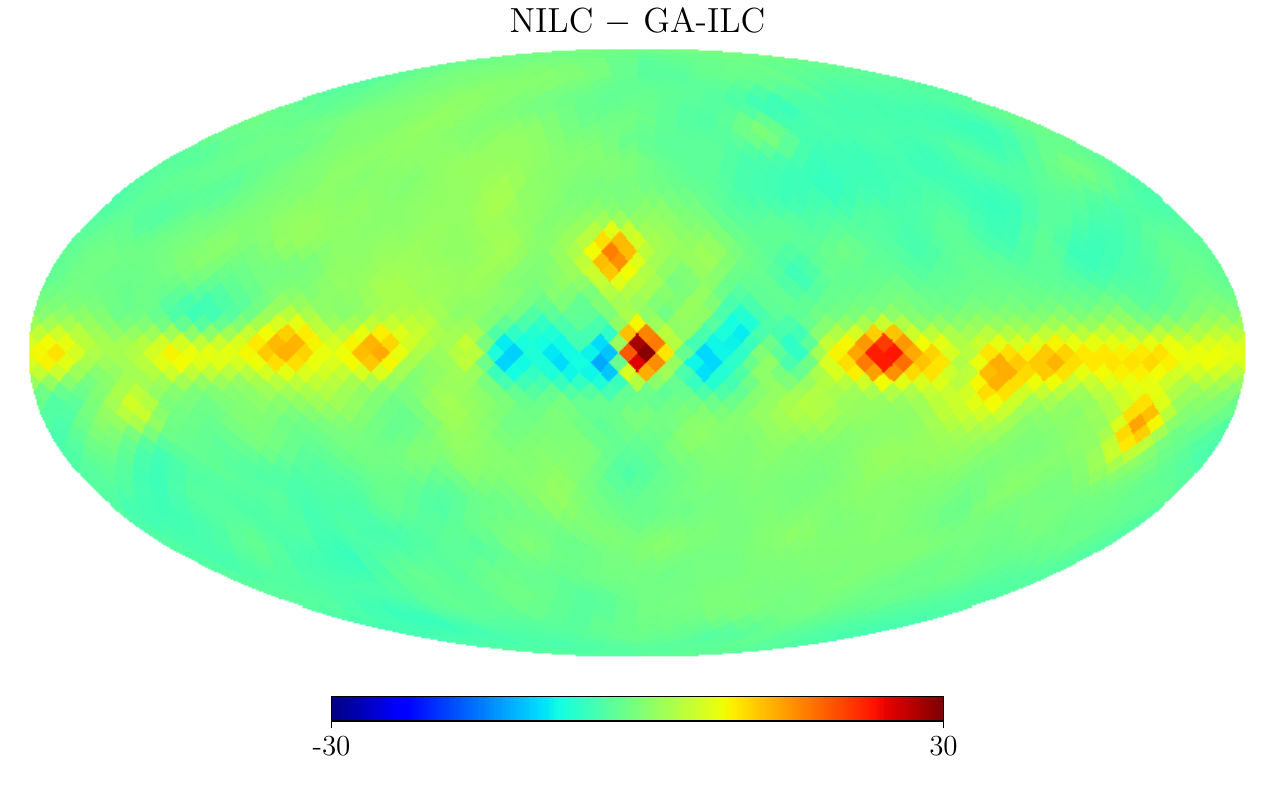}
	\includegraphics[width=0.49\linewidth]{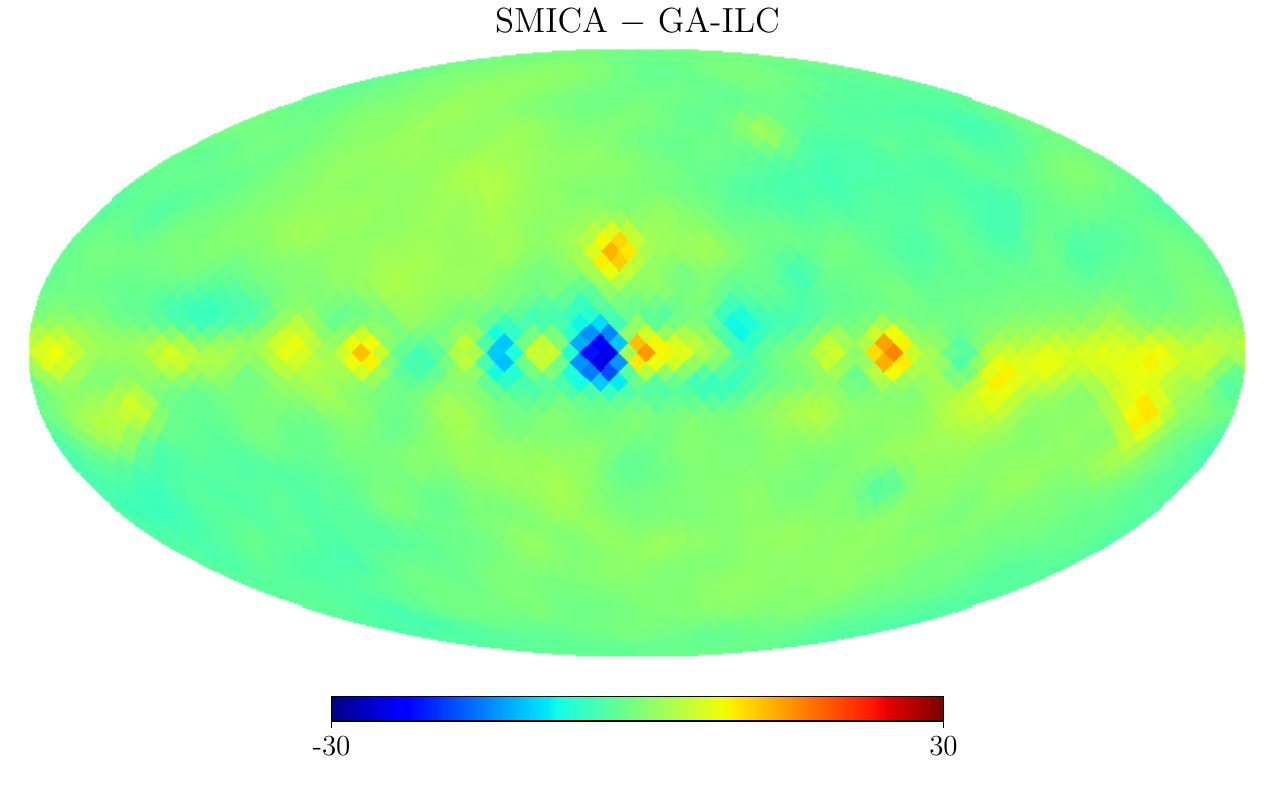}
	\includegraphics[width=0.49\linewidth]{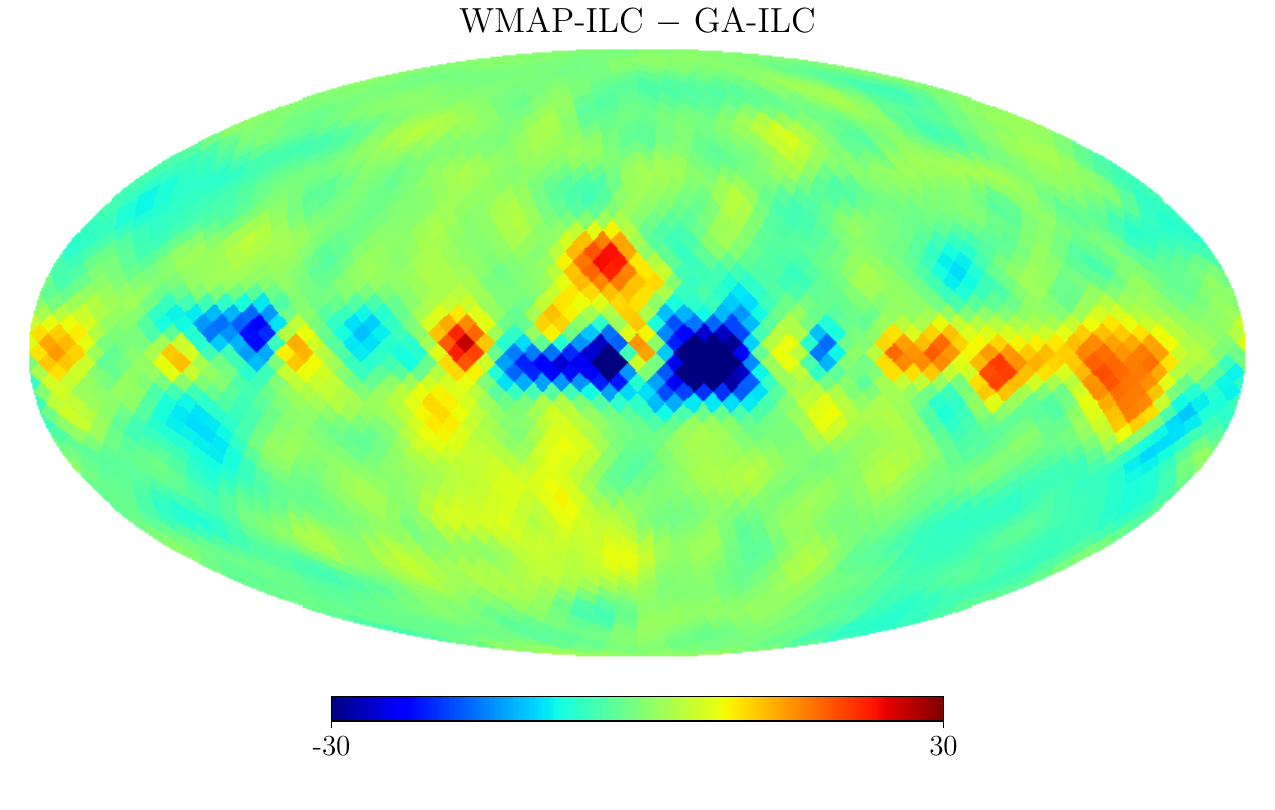}
	\caption{Difference maps of several other foreground reconstruction techniques and GA-ILC implemented on data. Notice that except the small residual foreground close to the galactic equator -- which itself varies from one method to the other, our results agree very nicely with those of other methods. The values are in $\mu$K.}
	\label{fig:comp_diff_maps}
\end{figure*}

Listed in table~\ref{tab:weights-SS-GA} are the weights produced by the analytical SS-ILC method and our GA-ILC method. We observe that the weights are significantly different in both the methods, and yet, they produce agreeable results. This asserts that our purely numerical GA-ILC method is as good as its analytical equivalent. This is a crucial takeaway of this work, in that the numerical method presents a reliable solution when an analytical approach is difficult or impossible.

\begin{figure*}
	\includegraphics[width = \linewidth]{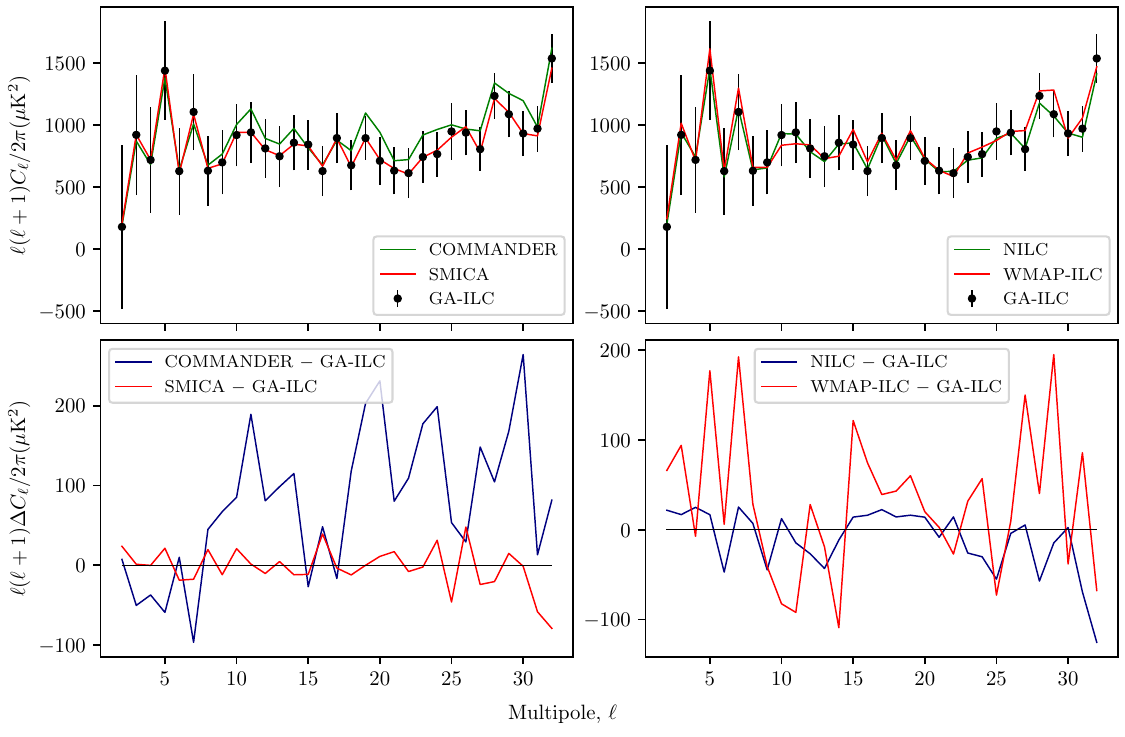}
	\caption{Comparison of angular power spectrum of the GA-ILC clean map with those of several different foreground reconstruction techniques. Bottom panel: differences of those various clean angular power spectra and GA-ILC clean power spectrum (data implementation). A good agreement between GA-ILC and other methods is seen here as well.}
	\label{fig:cl_comp}
\end{figure*}

\begin{table}
	\centering
	\caption{Reduced variance ($\sigma^2$) values of clean maps produced by different reconstruction methods for comparison.} \label{tab:red-var-comp}
	\begin{tabular}{|c|r|}
		\hline
		Method 				&	 $\sigma^2$ ($\mu$K$^2$) \\
		\hline
		SS-ILC				 & 	1003.34 \\
		GA-ILC				& 	1010.20 \\
		WMAP-ILC		& 	1042.55 \\
		SMICA				&  1003.35 \\
		NILC			 	 &	  990.47 \\
		COMMANDER   & 	1127.00 \\
		\hline
	\end{tabular}
\end{table}

We compare our GA-ILC results with those obtained by various component reconstruction methods of other science groups, namely, COMMANDER, NILC, SMICA \citep{Planck2018IV}, and WMAP-ILC \citep{WMAP-9yr-maps}. Shown in fig.~\ref{fig:comp_diff_maps} are the difference maps of each of those clean maps and the GA-ILC clean map (data implementation, of course). It is evident from here that our method produces a clean map that agrees well with some other completely different methods. We also observe that the residual contamination near the Milky Way plane is different in different methods. This indicates that those other clean maps also contain some (minimal) amount of residual foreground, same as ours. A comparison of the reduced variance values of the clean maps produced by different component-reconstruction methods is made in table~\ref{tab:red-var-comp}. It can be seen that, despite the stochastic nature of our GA-ILC method, it produces comparable results to the previous works. In fig.~\ref{fig:cl_comp} we present the angular power spectra of GA-ILC and that of COMMANDER, NILC, SMICA, and WMAP-ILC for comparison. As expected, the spectra agree well with each other, with the small difference between all the different methods. Again, the other methods' results differ from each other same as our method's.

\section{Conclusion}\label{sec:conclusion}

In this paper, for the first time in literature, we develop and implement the biological-selection-rule-motivated genetic algorithm to reconstruct the CMB component over large angular scales by removing foregrounds using linear combination of multifrequency observations of WMAP and Planck satellite missions. The genetic algorithm is a computational method that imitates the biological process of ``evolution by natural selection'' to find global optimal solutions (to multivariable problems) over generations of synthetic solution-sets called populations. A selection pressure is applied to individual solutions in terms of their ``fitness'' while reproducing a daughter generation by crossovers and mutation, the latter being an important factor for preserving genetic diversity. We perform a detailed study of the effects of mutation by running 128 GA-ILC instances with distinct but closely spaced mutation coefficients. We run each individual GA-ILC instance on a single Intel E5-2670 CPU core with 4 GB memory within a HPC cluster. We estimate the average time of execution of a GA-ILC instance to be 21 minutes. We validate our methodology by performing 200 Monte Carlo simulations using realistic observations from the final-year WMAP and Planck missions. The results of the simulations show that the CMB map and the angular power spectrum can be accurately recovered by using our method. The outcome of our method is in close agreement with the results obtained by using weights following the exact analytical method which demonstrates usefulness of the new method of this paper. We note that the reduced variance obtained by the GA-ILC is marginally higher than the SS-ILC case. The GA-ILC weights obtained by us are different from the SS-ILC weights. A future article will explore whether the GA-ILC weights may further be tuned to obtain reduced variance even closer to or lower than the SS-ILC case. We also compare the cleaned CMB maps and recovered angular power spectrum obtained by the new method with those obtained by WMAP and Planck science groups. These results agree well with each other, which shows that the CMB results obtained by the satellite missions are robust with respect to the data analysis algorithms applied. 

In a usual ILC framework (e.g., SS-ILC), one tries to minimize the two-point correlation function, namely the variance of the linearly combined clean map, but a mere variance minimization is non-optimal except in the case if the foregrounds are Gaussian random fields, since for non Gaussian distributions independent information can be obtained from higher order moments beyond variance. GA-ILC opens up an avenue to explore the general class of cost functions that incorporate information from such higher order moments obtained from the observed CMB maps as the foregrounds are strongly non-Gaussian. In this first implementation of GA-ILC we report that the method produces results that are competent with the analytical ILC methods. In a future work we aim to build on this to explore the potential of GA-ILC at its fullest. GA-ILC would be an excellent tool for application for physically well motivated cost-functions that do not possess any analytical solutions for weights at their global optima.

We note that natural selection is a slow and resource-demanding process, and this characteristic is also carried by our synthetic GA environment since it is designed to mimic the slow natural evolution. Nevertheless, the robust nature (\emph{w.r.t.} fluctuations in the optimizing system) of the global optimal solutions derived by the GA-ILC as demonstrated in this work, makes it the most interesting and appealing method (of all the multivariable local optimization algorithms in literature) in the context of CMB reconstruction. Enriched in inherent parametric characteristics, GA-ILC is a very flexible method. With the help of the control parameters like number of individuals, mutation probability, and the number of generations the method self-guidedly evolves towards the global optimal point, gradually producing fitter solutions at every successive generation. The larger domain of application of the GA-ILC is a very promising feature and will be explored in the future communications by the authors.

\section*{Acknowledgements}
We thank the HPC cluster facility,  ``Kanad'' of IISER Bhopal for providing access to its resources for a crucial part of this research.

\section{Data Availability}
The data pertaining to this article will be shared on reasonable request to the corresponding author.


\bibliographystyle{mnras}
\bibliography{main} % if your bibtex file is called example.bib

%\begin{thebibliography}{99}
%\end{thebibliography}

%%%%%%%%%%%%%%%%%%%%%%%%%%%%%%%%%%%%%%%%%%%%%%%%%%

% Don't change these lines
\bsp	% typesetting comment
\label{lastpage}
\end{document}